\documentclass[12pt,epsf,amssymb,amsmath,amsfonts]{article}
\usepackage{tabularx}
\usepackage{array}
\usepackage{graphics}
\usepackage{graphicx}
\usepackage{epsfig}
\usepackage{amsfonts}

\makeatletter

\usepackage{verbatim}

\newcommand{\msbar}{{\overline{\rm MS}}}
\newcommand{\ri}{{\rm RI/MOM}}

\newcommand{\bea}{\begin{eqnarray}}
\newcommand{\eea}{\end{eqnarray}}
\newcommand{\beq}{\begin{equation}}
\newcommand{\eeq}{\end{equation}}
\newcommand{\gev}{{\rm GeV}}

\newcommand{\ga}{{\gamma}}

\newcommand{\pdir}{p\kern -5.2pt\raise 0.2ex\hbox {/}}
\newcommand{\vdir}{v\kern -5.75pt\raise 0.15ex\hbox {/}}
\newcommand{\kdir}{k\kern -5.75pt\raise 0.15ex\hbox {/}}
\newcommand{\epsdir}{\epsilon\kern -5.0pt\raise 0.15ex\hbox {/}}
\newcommand{\bvdir}{\bar{v}\kern -5.75pt\raise 0.15ex\hbox {/}}
\newcommand{\Ddir}{D\kern -7.75pt\raise 0.20ex\hbox {/}}
\newcommand{\ldir}{l\kern -5.0pt\raise 0.2ex\hbox{/}}
\newcommand{\varepsdir}{\varepsilon\kern -5.5pt\raise 0.15ex\hbox{/}}

\newcommand{\bbar}{B^0-{\overline{B^0}}}

\makeatother

\begin{document}

\thispagestyle{empty}
\begin{flushright}
\begin{tabular}{l}
{\tt  IFUP-TH 2001/32}\\
{\tt  FTUV-IFIC-01-0927}\\
{\tt ROMA-1323/01}\\                                              
\end{tabular}
\end{flushright}
\vskip 2.2cm\par
\begin{center}
{\par\centering \textbf{\LARGE $B$-parameters of the complete} }\\
\vskip .45cm\par
{\par\centering \textbf{\LARGE set of matrix elements of $\Delta B = 2$} }\\
\vskip .45cm\par
{\par\centering \textbf{\LARGE operators from the lattice } }\\
\vskip 1.85cm\par
{\par\centering \large
\sc D.~Be\'cirevi\'c$^a$, V.~Gim\'enez$^b$, G.~Martinelli$^a$,\\
M.~Papinutto$^c$,  J.~Reyes$^{a,b}$}
\end{center}
{\par\centering \textsl{$^a$Dip.~di Fisica, Univ.~di Roma ``La
Sapienza''
and}\\
\textsl{INFN, Sezione di Roma, P.le A.~Moro 2, I-00185 Roma,
Italy.}\\
\vskip 0.3cm\par} 
{\par\centering \textsl{$^b$Dep.~de F\'\i sica Te\`orica and IFIC, Univ.~de
Val\`encia,}\\
\textsl{Dr.~Moliner 50, E-46100, Burjassot, Val\`encia,
Spain}\textsf{\textit{.}}\\
\vskip 0.3cm\par }                                                               
{\par\centering \textsl{$^c$Dip.~di Fisica, Univ.~di Pisa and INFN,
Sezione di Pisa,
}\\
\textsl{Via Buonarroti~2, I-56100 Pisa, Italy}\textsf{\textit{.}}\\
\vskip 1.1cm\par }                                                               
\begin{abstract}
We compute on the lattice the ``bag'' parameters of the five 
$\Delta B = 2$ operators of the supersymmetric basis, by combining 
their values determined in full QCD and in the static limit of HQET. 
The extrapolation of the QCD results 
from the accessible heavy-light meson masses to the $B$-meson mass is 
constrained by the static result. The matching of the corresponding 
results in HQET and in QCD is for the first time made at NLO accuracy 
in the $\msbar$(NDR) renormalization scheme. All results are obtained 
in the quenched approximation. 
\end{abstract}
\vskip 0.9cm
{\footnotesize {\bf PACS:} \sf 11.15.Ha (Lattice gauge theory), 
\ 12.38.Gc (Lattice QCD calculations),
\ 13.75Lb  (Meson-meson interactions),
\ 14.40.Nd   (Bottom mesons),
\  12.39.Hg (Heavy quark effective theory)}                                                    
\newpage

\setcounter{page}{1}
\setcounter{footnote}{0}
\setcounter{equation}{0}                                                        

\noindent
 
\renewcommand{\thefootnote}{\arabic{footnote}}
 
\setcounter{page}{1}
\setcounter{footnote}{0}
\setcounter{equation}{0}
\section{Introduction}

This paper is devoted to a combined analysis of the   
matrix elements of the complete set of  $\Delta B=2$ operators, which we computed 
on the lattice in both  
the static limit of the heavy quark effective theory (HQET)  
and in  standard lattice QCD (with Wilson fermions). All five 
operators enter the phenomenological analyses of supersymmetric (SUSY) 
effects that might affect the Standard Model (SM) expectations for 
$\Delta m_{B_d}$ and/or $\Delta m_{B_s}$. It is therefore convenient 
to work in the so-called SUSY basis of operators:
\bea
 \label{basisc}
{\phantom{\large{l}}}\raisebox{-.16cm}{\phantom{\large{j}}}
O_1 &=& \ \bar b^i \gamma_\mu (1- \gamma_{5} )  q^i \,
 \bar b^j  \gamma_\mu (1- \gamma_{5} ) q^j \,  , 
  \nonumber \\
{\phantom{\large{l}}}\raisebox{-.16cm}{\phantom{\large{j}}}
O_2 &=& \ \bar b^i  (1- \gamma_{5} ) q^i \,
\bar b^j  (1 - \gamma_{5} )  q^j \, ,  \nonumber  \\
{\phantom{\large{l}}}\raisebox{-.16cm}{\phantom{\large{j}}}
O_3&=& \ \bar b^i  (1- \gamma_{5} ) q^j \,
 \bar b^j (1 -  \gamma_{5} ) q^i \, ,  \\
{\phantom{\large{l}}}\raisebox{-.16cm}{\phantom{\large{j}}}
O_4 &=& \ \bar b^i  (1- \gamma_{5} )  q^i \,
 \bar b^j   (1+ \gamma_{5} ) q^j \,  ,  \nonumber \\
{\phantom{\large{l}}}\raisebox{-.16cm}{\phantom{\large{j}}}
O_5 &=& \ \bar b^i  (1- \gamma_{5} )  q^j \,
 \bar b^j   (1+ \gamma_{5} ) q^i \,  ,  \nonumber 
 \eea                                                                            
where the superscripts denote colour indices, and $q$ stands for 
either $d$- or $s$- light quark flavour. 
The first of the above operators has been widely studied over the last decade,  
since it is crucial for the SM description of the $\bbar$ mixing
amplitude, whereas 
$O_2$ and $O_3$ were also recently studied because they are relevant for the  
SM estimates of the relative width difference in the neutral $B$-meson system, 
$\left( \Delta \Gamma/\Gamma\right)_{B_s}$~\cite{Bernard:2001ki}.

It is customary to parameterize the matrix elements of the operators~(\ref{basisc}) 
in terms of the so-called ``bag''-parameters, which are introduced as a 
measure of the mismatch between the vacuum saturation approximation (VSA) 
and the actual value for each of the matrix elements, namely~\cite{luca,allton}
\bea
 \label{params}
\langle \bar B^0_q \vert \hat O_1(\mu) \vert   B^0_q \rangle  &=& {8\over 3} \, m_{B_q}^2  f_{B_q}^2
B_1(\mu) \ ,  \nonumber \\
\langle \bar B^0_q \vert \hat O_2(\mu) \vert   B^0_q \rangle  &=& -{5\over 3} \, \left( {m_{B_q}\over
m_b(\mu) + m_q(\mu) } \right)^2
m_{B_q}^2  f_{B_q}^2
B_2(\mu) \, ,  \nonumber  \\
\langle \bar B^0_q \vert \hat O_3(\mu) \vert   B^0_q \rangle &=& {1\over 3} \, \left( {m_{B_q}\over
m_b(\mu) + m_q(\mu) } \right)^2
m_{B_q}^2  f_{B_q}^2
B_3(\mu)\, ,  \\
\langle \bar B^0_q \vert \hat O_4 (\mu) \vert   B^0_q \rangle &=&  2 \, \left( {m_{B_q}\over
m_b(\mu) + m_q(\mu) } \right)^2
m_{B_q}^2  f_{B_q}^2
B_4(\mu)\,  ,  \nonumber \\
\langle \bar B^0_q \vert \hat O_5(\mu) \vert   B^0_q \rangle  &=& {2\over 3} \, \left( {m_{B_q}\over
m_b(\mu) + m_q(\mu) } \right)^2
m_{B_q}^2  f_{B_q}^2
B_5(\mu)\,  .  \nonumber 
 \eea                                                                            
The hat symbol denotes operators renormalized in some 
renormalization scheme at the renormalization scale $\mu$. 
To determine the values of the ``bag'' parameters $B_{1-5}(\mu)$, we
have performed a numerical simulation 
of (quenched) QCD on the lattice. 
While such a simulation can be made directly for the $c$-quark mass or 
somewhat heavier, present limitations of computational resources  
do not allow for a direct study of the $b$-quark.
For this reason we work in the range of heavy-light pseudoscalar 
masses $m_P\in (1.7, 2.4)$~GeV, from which we have to 
extrapolate to the physical point, $m_{B_d}=5.28$~GeV and/or 
$m_{B_s}=5.37$~GeV. 
Although guided by the HQET scaling laws, this extrapolation 
is the dominant source of the systematic uncertainty in final results. 
To get around this problem we also computed the same 
matrix elements in the static limit of HQET on the lattice, and used them to   
constrain the extrapolations towards the physical point, 
$m_{B_{s/d}}$. It is technically challenging to combine 
results from two different theories, namely one should match 
the $B$-parameters obtained in QCD onto the HQET ones so that 
the heavy quark scaling laws can be safely used. 
A special care to this issue will be given in the body of this paper.

The main features of this work are:
\begin{itemize}
\item[--] The $B$-parameters which appear in eq.~(\ref{params}) are computed 
using lattice QCD with Wilson fermions and are renormalized 
non-perturbatively in the (Landau) RI/MOM scheme. It is important to 
stress that we incorporated the recent proposal to remove the effects 
of the Goldstone boson contamination~\cite{alain,dawson};
\item[--] The $B$-parameters computed in  the static limit of HQET on the lattice 
are matched onto the continuum $\msbar$(NDR) renormalization 
scheme. This matching has been made by using the one-loop (boosted) perturbative 
expressions~\cite{GR1,massimo};
\item[--] The conversion of the operators computed in lattice QCD from RI/MOM 
to $\msbar$(NDR) scheme is made at NLO accuracy~\cite{ciuchini,buras}. 
 Matching of the QCD operators onto the HQET ones has also been 
performed at NLO accuracy in a specified $\msbar$(NDR) scheme. In that procedure
we use the recently computed 2-loop anomalous dimension matrices in HQET~\cite{GR2}. 
With this matching at hand, we were able to 
constrain the extrapolation, {\it i.e.} to interpolate to the 
physical $b$-quark mass;
\item[--] Final results are presented in the RI/MOM scheme and in the 
$\msbar$(NDR) scheme of ref.~\cite{buras}. 
In addition, the parameters $B_{1,2,3}$ are also given in the $\msbar$(NDR) 
renormalization scheme of ref.~\cite{beneke}.~\footnote{Preliminary results were presented in
ref.~\cite{juan}.}
\end{itemize}
The complete list of results can be found in table~\ref{tab0}. Notice that we  
do not observe any $SU(3)$ breaking effect in the $B$-parameters, {\it i.e.}:
\bea
\left. {\, B_i^{(s)} \over B_i^{(d)} } \; \right|_{i=1,\dots,5} \ =\ 
\biggl\{  0.99(2),\ 1.01(2),\ 1.01(3),\ 1.01(2),\ 
1.01(3) \biggr\} \ .
\eea
\begin{table}[t!] 
\begin{center} 
\begin{tabular}{|c|c|c|c|} 
\hline
{\phantom{\Huge{l}}}\raisebox{-.2cm}{\phantom{\Huge{j}}}
\hspace*{-5mm}{\sl Scheme}&  RI/MOM  &  $\msbar$(NDR)~\cite{buras}  &  $\msbar$(NDR)~\cite{beneke} 
  \\   \hline \hline 
{\phantom{\Huge{l}}}\raisebox{-.2cm}{\phantom{\Huge{j}}}  
\hspace*{-5mm} $B_1^{(d)}(m_b)$ & $0.87(4)\left({}^{+5}_{-4}\right)$ & $0.87(4)\left({}^{+5}_{-4}\right)$
&   $0.87(4)\left({}^{+5}_{-4}\right)$ \\ 
{\phantom{\Huge{l}}}\raisebox{-.2cm}{\phantom{\Huge{j}}}
\hspace*{-5mm} $B_2^{(d)}(m_b)$ & $0.82(3)(4)$ & $0.79(2)(4)$ &  $0.83(3)(4)$\\ 
{\phantom{\Huge{l}}}\raisebox{-.2cm}{\phantom{\Huge{j}}}
\hspace*{-5mm} $B_3^{(d)}(m_b)$ & $1.02(6)(9)$ & $0.92(6)(8)$ &  $0.90(6)(8)$ \\ 
{\phantom{\Huge{l}}}\raisebox{-.2cm}{\phantom{\Huge{j}}}
\hspace*{-5mm} $B_4^{(d)}(m_b)$ & $1.16(3)\left({}^{+5}_{-7}\right)$ 
&  $1.15(3)\left({}^{+5}_{-7}\right)$ & -- \\ 
{\phantom{\Huge{l}}}\raisebox{-.2cm}{\phantom{\Huge{j}}}
\hspace*{-5mm} $B_5^{(d)}(m_b)$ & $1.91(4)\left({}^{+22}_{-7}\right)$ & $1.72(4)\left({}^{+20}_{-6}\right)$&  --   
  \\ \hline
{\phantom{\Huge{l}}}\raisebox{-.2cm}{\phantom{\Huge{j}}}  
\hspace*{-5mm} $B_1^{(s)}(m_b)$ & $0.86(2)\left({}^{+5}_{-4}\right)$ & 
$0.87(2)\left({}^{+5}_{-4}\right)$
&   $0.87(2)\left({}^{+5}_{-4}\right)$ \\ 
{\phantom{\Huge{l}}}\raisebox{-.2cm}{\phantom{\Huge{j}}}
\hspace*{-5mm} $B_2^{(s)}(m_b)$ & $0.83(2)(4)$ & $0.80(1)(4)$ &  $0.84(2)(4)$\\ 
{\phantom{\Huge{l}}}\raisebox{-.2cm}{\phantom{\Huge{j}}}
\hspace*{-5mm} $B_3^{(s)}(m_b)$ & $1.03(4)(9)$ & $0.93(3)(8)$ &  $0.91(3)(8)$ \\ 
{\phantom{\Huge{l}}}\raisebox{-.2cm}{\phantom{\Huge{j}}}
\hspace*{-5mm} $B_4^{(s)}(m_b)$ & $1.17(2)\left({}^{+5}_{-7}\right)$ &  $1.16(2)\left({}^{+5}_{-7}\right)$ & -- \\ 
{\phantom{\Huge{l}}}\raisebox{-.2cm}{\phantom{\Huge{j}}}
\hspace*{-5mm} $B_5^{(s)}(m_b)$ & $1.94(3)\left({}^{+23}_{-7}\right)$ & $1.75(3)\left({}^{+21}_{-6}\right)$&  --   
  \\ \hline
\end{tabular} 
\caption{\label{tab0}
\small{\sl The main results of this paper: $B$-parameters defined in
eq.~(\ref{params}), renormalized at $\mu = m_b = 4.6$~GeV and in three 
renormalization schemes: $\ri$, $\msbar$ of ref.~\cite{buras} and the
$\msbar$ of ref.~\cite{beneke}. The results are obtained in the quenched
approximation.}}
\end{center}
\vspace*{-.3cm}
\end{table}
This paper is organized as follows: In sec.~\ref{sec2} 
we give the essential details of our lattice calculations and
present the results as obtained for each heavy quark that we were able to access
from our lattice and also in the static limit of the HQET. In sec.~\ref{sec3} we 
outline the general strategy to combine the results of the two theories. We then 
explicitly give all the necessary anomalous dimension matrices and present the  
results of the combined analysis for all the five $B$-parameters. 
In sec.~\ref{sec4} we discuss the systematic uncertainties which are included 
in the results given in table~\ref{tab0}. We briefly conclude in sec.~\ref{sec5}.

\section{Direct Lattice results\label{sec2}}

\subsection{Computation in lattice QCD}
In this subsection we recall the main elements of our lattice 
simulation, the details of which can be found in refs.~\cite{ape1,ape2}. 
We work with a lattice of the size $24^3\times 48$, at $\beta = 6.2$, and 
use the non-perturbatively improved Wilson action~\cite{luscher0}. 
Note however that the 4-fermion operators, which are the main target of the present 
work, are not improved. Our data-set consists of 200 independent gauge field 
configurations. We work with 3 values of the heavy and 3 values of
the light quark masses, corresponding to the Wilson hopping 
parameters: $\kappa_q \in \{ 0.1344, 0,1349, 0.1352\}$, and 
$\kappa_Q \in \{ 0.125, 0,122, 0.119\}$. The mass spectrum and 
the decay constants have already been discussed in our previous 
publications~\cite{ape1,ape2,ape3} and we immediately turn to the 
computation of the $B$-parameters.

The starting point is to compute the 2- and 3-point correlation functions
\bea \label{23fns}
&&{\cal C}_{{JJ}}^{(2)}(t) = \langle \displaystyle{\sum_{\vec x}} J({\vec x}, t)
J^{\dagger}(0) \rangle\, \stackrel{t\gg 0}{\longrightarrow}\, 
{{\cal{Z}}_J \over  2 \sinh M_J} \, e^{- M_J
t} \; ,\nonumber \\
&&{\cal C}^{(3)}_i (t_{1}, t_{2}) =
  \langle \displaystyle{\sum_{\vec x,\vec y}} P_5 (\vec x, t_{2})
  \hat O_i(\vec 0, 0;\mu)  P_5^\dagger (\vec y, t_{1}) \rangle \, 
   \nonumber \\
&&\hspace*{2.2cm}\stackrel{T-t_2\gg 0}{\longrightarrow}\, 
{\sqrt{{\cal{Z}}_P} \over  2 \sinh M_P} e^{- M_P t_1}\cdot \langle P_q\vert \hat O_i(\mu)\vert  P_q\rangle 
\cdot {\sqrt{{\cal{Z}}_P} \over  2 \sinh M_P} e^{- M_P t_2}  \  ,
\eea
where the operator $\hat O_i$ is placed at time equal to zero, the source of 
the pseudoscalar mesons ($P_5=\bar Q \gamma_5 q$) is fixed at $t_1=16$,
while the other source operator moves around the periodic lattice 
(of the size $T=48$). At some $t_2\equiv t$, which is sufficiently far 
from  the first source and from the operator $\hat O_i$, the 
lowest lying heavy-light pseudoscalar meson, $P_q$, is isolated. $J$ in the
above equations stands for either $P_5$ or $A_0=\bar Q \gamma_0 \gamma_5 q$.

To extract the parameter $B_1(\mu)$, one computes
\bea \label{eq1}
&& R_{B_1}(t)={ {\cal C}^{(3)}_1 (t_{1}, t; \mu) \over \  \displaystyle{8\over 3}
 Z_A^2 \ {\cal C}_{AP}^{(2)} 
(t) \ {\cal C}_{AP}^{(2)} (t_{1})}\ \stackrel{T-t\gg 0}{\longrightarrow}\,\ 
{\; \langle   P_q \vert \hat O_{1}(\mu)  \vert   P_q \rangle \; \over
 \displaystyle{8\over 3} \vert \langle 0 \vert \hat A_0 \vert   P_q \rangle \vert^2} \ \equiv B_1(\mu) \ ,
\eea
where the 2-points functions are used to eliminate the exponential terms 
from the 3-point functions~(\ref{23fns}) and also to divide out 
the $(8/3) f_P^2 m_P^2$ from eq.~(\ref{params}), thus
accessing directly the wanted $B$-parameter. Similarly, to reach other 
four $B$-parameters we form the ratios 
\bea \label{eq2}
&& R_{B_i}(t)={ {\cal C}^{(3)}_i (t_{1}, t; \mu) \over \  b_i\  
 Z_P^2(\mu)\ {\cal C}_{PP}^{(2)} 
(t) \ {\cal C}_{PP}^{(2)} (t_{1})}\ \stackrel{T-t\gg 0}{\longrightarrow}\,\ 
{\langle   P_q \vert \hat O_{i}(\mu)  \vert   P_q \rangle \over
 b_i \ \vert \langle 0 \vert \hat P_5(\mu) \vert   P_q \rangle \vert^2} \ \equiv B_i(\mu) \ ,
\eea
where $b_i \in \{ -5/3, 1/3,2,2/3 \}$ for $i\in \{2,3,4,5\}$, respectively.

As already discussed in our previous works~\cite{ape1,ape2}, 
the operators $\hat O_i(\mu)$ are renormalized non-perturbatively in 
the (Landau)RI/MOM scheme by using the method
explained in detail in ref.~\cite{npr} (see also references therein). 
The method is based on the possibility of computing amputated 4-quark 
vertices at sufficiently large quark virtualities with the 
operators inserted at zero momentum. 
The RI/MOM renormalization condition is merely imposed on various 
projections of the amputated Green functions, which then lead to a full 
set of $9$ renormalization ($Z(\mu, g_0^2)$) and $16$ subtraction 
($\Delta(g_0^2)$) constants, where $g_0^2 = 6/\beta$ is the bare lattice 
coupling. The renormalized operators 
are given by
\bea
\hat O_i(\mu) \ =  \ Z_{ij} (\mu, g_0^2) \ \left[  O_j(g_0^2) \ + \ \Delta_{jk} (g_0^2) O_k(g_0^2)
\right]_{k\neq j} \; .
\eea 
Since the computation of the off-shell quantities (4-quark vertices) is
performed in the Landau gauge, the scheme is often referred to as 
the Landau RI/MOM scheme.

In practice, one computes all the renormalization and subtraction constants 
with specific (nonzero) values of the light quark masses and then extrapolates 
all $Z$'s and $\Delta$'s to the chiral limit. It has been pointed out in 
refs.~\cite{alain,dawson} that such an extrapolation can be contaminated 
by the Goldstone boson (GB) contributions. 
The recipe to subtract these contributions away has been proposed 
and implemented in ref.~\cite{DI32} (see also appendix of this paper). 
We employed that prescription and recomputed the renormalization and
subtraction constants. Their values, together 
with $Z_P(\mu)$ and $Z_A$ which 
are needed in eqs.~(\ref{eq1},\ref{eq2}),  
are listed in appendix. 
This is a new feature of our computation which improves (corrects) our 
previous results, presented in refs.~\cite{ape1,ape2}.

\begin{figure}[h!]
\begin{center}
\begin{tabular}{c c c}
\hspace*{-11mm}&\epsfxsize17.3cm\epsffile{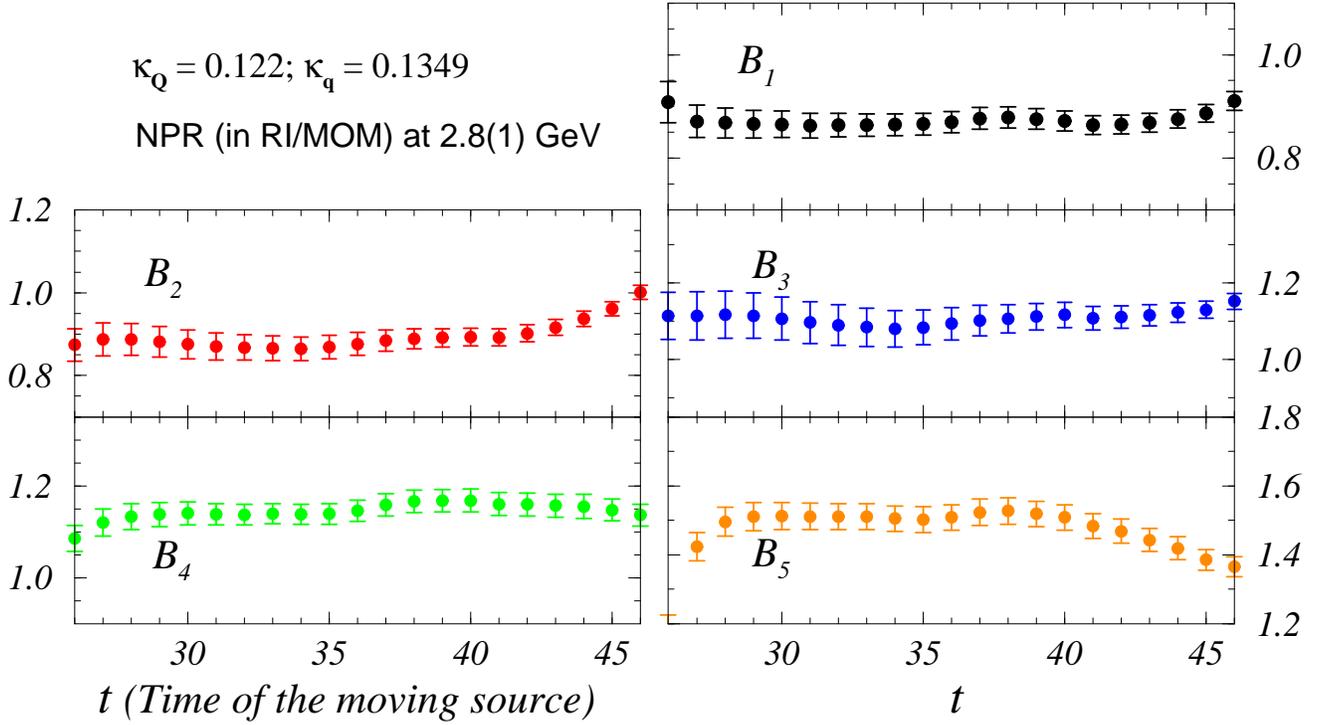} &  \\
\end{tabular}

\caption{\label{fig1}{\sl \small Signals for the ratios $R_{B_i}(t)$, 
defined in eqs.~(\ref{eq1}) and (\ref{eq2}), are shown for the combination:
$\kappa_q=0.1349$, $\kappa_Q=0.122$. The operators are non-perturbatively 
renormalized (NPR) in the (Landau)RI/MOM scheme at $\mu \simeq 1/a
=2.8(1)$~GeV.}}
\end{center}
\end{figure}

\begin{figure}[h!]
\begin{center}
\begin{tabular}{c c c}
\hspace*{-11mm}&\epsfxsize17.3cm\epsffile{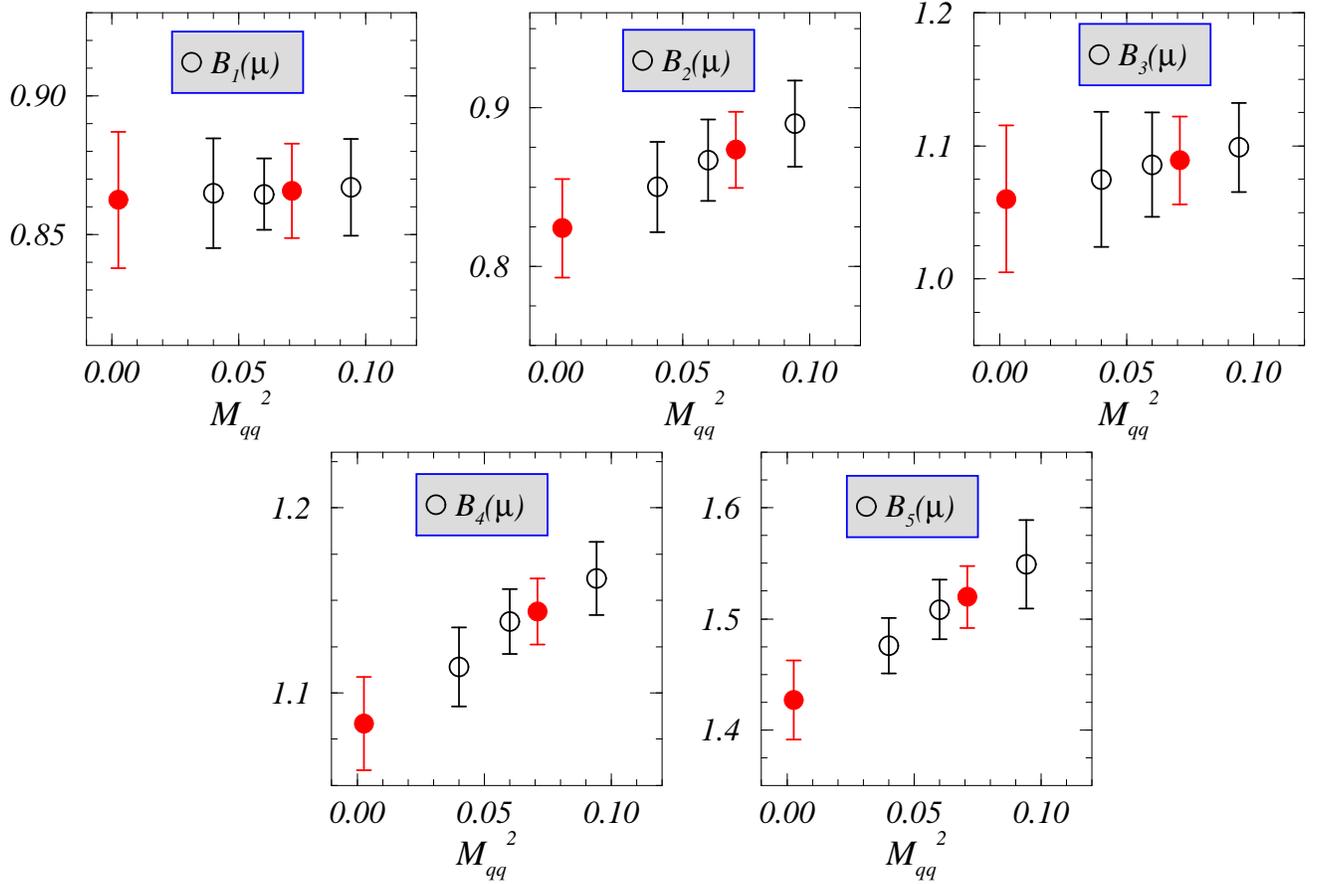} &  \\
\end{tabular}

\caption{\label{fig2}{\sl \small Extrapolations to the light up/down 
and interpolations to
the strange light quark mass are shown for all five $B$-parameters 
in the case of fixed heavy quark mass corresponding to $\kappa_Q=0.122$. 
Empty symbols denote 
the $B$-parameters directly accessed in our simulation. Filled 
symbols correspond to the $B$ parameters extrapolated to the physical 
$u/d$ and $s$ quark respectively. 
In the figure $\mu a = 1.03$.}}
\end{center}
\end{figure}

In fig.~\ref{fig1}, we illustrate the quality of the signals for 
the ratios $R_{B_{1-5}}(t)$ for a specific combination of 
heavy and light quarks.
After inspecting the ratios for all 9 pairs of $\kappa_Q$--$\kappa_q$, 
we find that common stability plateaus are reached for
\bea
R_{B_1}&:& t\in [28,35]  \,;\nonumber \\  
R_{B_{2,3}}&:& t\in [30,35]   \,;\nonumber \\
R_{B_{4,5}}&:& t\in [29,35]    \,.  
\eea
On each of these plateaus we fit the ratios to a constant and hence extract the
parameters $B_{1-5}(\mu)$ for each combination of the heavy and the light 
quark masses.
Every parameter is then linearly interpolated in the light quark mass to 
the {\em strange} and extrapolated to the {\em up/down} light quark mass, 
by using the (standard) 
lattice-plane method~(see refs.~\cite{oldape} for details). 
The extrapolations are very smooth as it can be seen from fig.~\ref{fig2}.
In table~\ref{tab2}, we present a detailed list of the results for all five 
$B$-parameters renormalized non-perturbatively at three different values of the 
renormalization scale $\mu$ in the (Landau)RI/MOM scheme. The results are
given for the values of 
the heavy quark masses, which are in the region of the charm
quark and slightly higher, and for the light quark interpolated to the 
strange and to the averaged up/down mass.
\begin{table}[h!] 
\begin{center} 
\begin{tabular}{|c|c|c|c||c|c|c|} 
\cline{2-7}
\multicolumn{1}{l|}{}&\multicolumn{3}{c||}
{\sl \small q=up/down}& 
\multicolumn{3}{c|}{\sl \small q=strange} \\
\hline 
{\phantom{\Huge{l}}}\raisebox{-.2cm}{\phantom{\Huge{j}}}
\hspace*{-7mm}{\sl Scale ($\mu$)}  & {1.9(1) GeV} & 2.8(1) GeV   &  3.9(2) GeV 
&  1.9(1) GeV & 2.8(1) GeV   &  3.9(2) GeV 
          \\ \hline \hline 
{\phantom{\Huge{l}}}\raisebox{-.2cm}{\phantom{\Huge{j}}}  
\hspace*{-5mm} $B_1^{(Q_1)}(\mu)$ & 
0.856(20) & 0.841(20) & 0.825(20) & 0.866(15) &0.850(15) & 0.835(15)\\ 
{\phantom{\Huge{l}}}\raisebox{-.2cm}{\phantom{\Huge{j}}}
\hspace*{-5mm} $B_2^{(Q_1)}(\mu)$ & 
 0.875(38) & 0.817(31)&  0.797(31) &0.904(27) &0.843(22)& 0.822(22)\\ 
{\phantom{\Huge{l}}}\raisebox{-.2cm}{\phantom{\Huge{j}}}
\hspace*{-5mm} $B_3^{(Q_1)}(\mu)$ & 
1.295(73) & 1.072(51) &  0.941(42)& 1.284(48)  &1.074(34) & 0.947(28)\\
{\phantom{\Huge{l}}}\raisebox{-.2cm}{\phantom{\Huge{j}}}
\hspace*{-5mm} $B_4^{(Q_1)}(\mu)$ & 
1.092(23) & 1.107(22) &  1.035(20)&  1.112(16) &1.129(16) &  1.056(15)\\
{\phantom{\Huge{l}}}\raisebox{-.2cm}{\phantom{\Huge{j}}}
\hspace*{-5mm} $B_5^{(Q_1)}(\mu)$ & 
1.523(37) & 1.322(29) &  1.332(27)&  1.606(30) &1.386(24) & 1.392(22)\\ \hline
{\phantom{\Huge{l}}}\raisebox{-.2cm}{\phantom{\Huge{j}}}  
\hspace*{-5mm} $B_1^{(Q_2)}(\mu)$ & 
0.880(25) & 0.862(24) & 0.849(24) &  0.883(17) &0.866(17) &  0.852(17)\\ 
{\phantom{\Huge{l}}}\raisebox{-.2cm}{\phantom{\Huge{j}}}
\hspace*{-5mm} $B_2^{(Q_2)}(\mu)$ & 
0.886(37) &  0.824(31)&  0.803(30) &0.940(26) & 0.874(22)&  0.851(22)\\ 
{\phantom{\Huge{l}}}\raisebox{-.2cm}{\phantom{\Huge{j}}}
\hspace*{-5mm} $B_3^{(Q_2)}(\mu)$ & 
1.275(81) & 1.060(55) &  0.932(45)&  1.294(48) &1.089(33) &   0.963(28)\\
{\phantom{\Huge{l}}}\raisebox{-.2cm}{\phantom{\Huge{j}}}
\hspace*{-5mm} $B_4^{(Q_2)}(\mu)$ & 
1.066(26) & 1.083(25) &  1.013(24)&  1.124(17) &1.144(17) &   1.070(16)\\
{\phantom{\Huge{l}}}\raisebox{-.2cm}{\phantom{\Huge{j}}}
\hspace*{-5mm} $B_5^{(Q_2)}(\mu)$ & 
1.673(45) & 1.427(36) & 1.421(34)&  1.784(35) &1.520(28) &  1.511(26)\\ \hline
{\phantom{\Huge{l}}}\raisebox{-.2cm}{\phantom{\Huge{j}}}  
\hspace*{-5mm} $B_1^{(Q_3)}(\mu)$ & 
0.887(26) & 0.869(25) & 0.856(25) &  0.886(16) &0.867(15) & 0.855(15)\\ 
{\phantom{\Huge{l}}}\raisebox{-.2cm}{\phantom{\Huge{j}}}
\hspace*{-5mm} $B_2^{(Q_3)}(\mu)$ & 
0.950(33) & 0.881(27)&  0.857(26) &0.962(23) & 0.890(19)&  0.869(19)\\ 
{\phantom{\Huge{l}}}\raisebox{-.2cm}{\phantom{\Huge{j}}}
\hspace*{-5mm} $B_3^{(Q_3)}(\mu)$ & 
1.349(84) & 1.123(54) & 0.989(43)&  1.288(51) & 1.090(34) & 0.966(27)\\
{\phantom{\Huge{l}}}\raisebox{-.2cm}{\phantom{\Huge{j}}}
\hspace*{-5mm} $B_4^{(Q_3)}(\mu)$ & 
1.114(23) & 1.136(21) & 1.062(20)&  1.119(14) & 1.141(13) &  1.068(12)\\
{\phantom{\Huge{l}}}\raisebox{-.2cm}{\phantom{\Huge{j}}}
\hspace*{-5mm} $B_5^{(Q_3)}(\mu)$ & 
1.885(47) & 1.592(35) &  1.573(33)&  1.925(32) & 1.622(25) & 1.601(23)\\ \hline
\end{tabular} 
\caption{\label{tab2}
\small{\sl $B$-parameters, as defined in eq.~(\ref{params}), extracted from our non-perturbatively 
renormalized data at three values of the renormalization scale $\mu$ in the RI/MOM 
renormalization scheme. 
From top to bottom of the table, the bag parameters correspond to the heavy quark with 
$\kappa_{Q_1}=0.125$, $\kappa_{Q_2}=0.122$ and $\kappa_{Q_3}=0.119$, respectively 
(they are separated by the horizontal lines). }}
\end{center}
\vspace*{-.3cm}
\end{table}
\subsection{Computation in the static limit of HQET}

To avoid a confusion in notations, we consistently use 
``tilde'' symbols over the operators and the $B$-parameters 
computed in HQET. 
Instead of the five operators that we listed in eq.~(\ref{basisc}), 
in HQET one deals with only four of them, namely
\bea
 \label{basish}
\widetilde O_1 &=& \ \bar h^i \gamma_\mu (1- \gamma_{5} )  q^i \,
 \bar h^j  \gamma_\mu (1- \gamma_{5} ) q^j \,  , 
  \nonumber \\
\widetilde O_2 &=& \ \bar h^i  (1- \gamma_{5} ) q^i \,
\bar h^j  (1 - \gamma_{5} )  q^j \, ,  \nonumber  \\
\widetilde O_4 &=& \ \bar h^i  (1- \gamma_{5} )  q^i \,
 \bar h^j   (1+ \gamma_{5} ) q^j \,  ,  \nonumber \\
\widetilde O_5 &=& \ \bar h^i  (1- \gamma_{5} )  q^j \,
 \bar h^j   (1+ \gamma_{5} ) q^i \,  ,   
 \eea                                                                            
where $h$ stands for the infinitely heavy (static) quark. In the HQET,  
the operator $\widetilde O_3$ is related to  
$\widetilde O_1$ and $\widetilde O_2$ by the equations of motion as
\bea \label{eq3}
\widetilde O_3 = - \widetilde O_2 - {1\over 2} \widetilde O_1\;.
\eea
The computation of the first two operators has been explained in detail 
in refs.~\cite{GR1,GRproc}.
The data-set consists of 600 configurations gathered on 
the $24^3 \times 40$ lattice at $\beta = 6.0$. The light quarks were 
simulated by using the tree-level improved Wilson action and three 
values of $\kappa_q \in\{ 0.1425, 1432, 1440\}$.

\begin{figure}[h!]
\vspace*{5mm}
\begin{center}
\begin{tabular}{c c c}
\hspace*{-11mm}&\epsfxsize17.3cm\epsffile{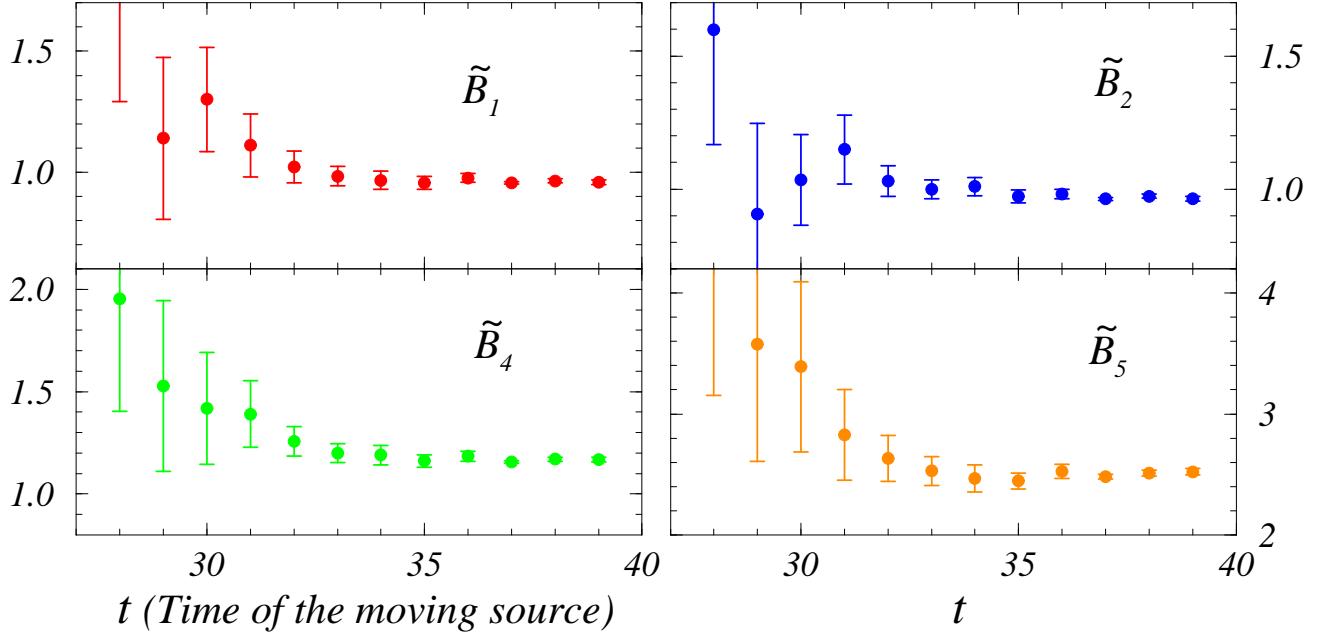} &  \\
\end{tabular}
\caption{\label{figH}{\sl \small  
Signals for the ratios $\widetilde R_{B_i}(t)$ 
defined in eqs.~(\ref{hqetR}) from which the $\widetilde B_i$ parameters are 
extracted. The common plateaus are chosen for $t_1\equiv t\in [32,36]$. 
The plotted ratios correspond to $\kappa_q=0.1432$. 
The operators are renormalized perturbatively and evolved to 
$\mu = m_b$.} }
\end{center}
\end{figure}

To reach the HQET parameters equivalent to the ones appearing in eq.~(\ref{params}), 
which we will call $\widetilde B_i$, one computes the following 2- and 3-point 
correlation functions:
\bea \label{hqet}
&&\widetilde {\cal C}^{(2)}_{AA}(t) = \langle \displaystyle{\sum_{\vec x}} \widetilde A_0({\vec x}, t)
\widetilde A_0^{\dagger}(0) \rangle\, \stackrel{t\gg 0}{\longrightarrow}\, {\widetilde  {\cal{Z}}_A} e^{-
\Delta {\cal E} t} \; ,\nonumber \\
&&\widetilde {\cal C}^{(3)}_i (t_{1}, t_{2}) =
  \langle \displaystyle{\sum_{\vec x,\vec y}}\widetilde  A_0 (\vec x, t_{2})
  \widetilde  O_i(\vec 0, 0;\mu) \widetilde  A_0^\dagger (\vec y, t_{1}) \rangle \, 
   \nonumber \\
&&\hspace*{2.2cm}\stackrel{t\gg 0}{\longrightarrow}\, 
\widetilde {\cal{Z}}_A  \cdot { \langle P_q\vert \widetilde O_i(\mu)\vert  P_q\rangle  \over  2  M_P}
\cdot e^{- \Delta {\cal E} (t_1-t_2)}  \  ,
\eea
where the source operators are the axial currents~\footnote{
In the static limit the axial current is
identical to the pseudoscalar density.} which are extended by using 
the so-called cubic smearing procedure~\cite{GM}. 
$\Delta {\cal E}$ in eq.~(\ref{hqet}), is the binding energy of the 
heavy meson $P_q$. 
Note also that in this case, $\sqrt{\widetilde {\cal{Z}}_A} = 
\langle 0\vert \widetilde A_0\vert P_q\rangle/\sqrt{2 M_P}$. 
To get the desired $B$-parameters, we form the following ratios:
\bea \label{hqetR}
&& \widetilde R_{B_i}(t_1)={ \widetilde {\cal C}^{(3)}_i (t_{1}, t_2; \mu) \over \  b_i\  
 Z_A^2\ \widetilde {\cal C}_{AA}^{(2)} 
(t_{2}) \ \widetilde {\cal C}_{AA}^{(2)} (t_{1})}\ \stackrel{t_1 \gg 0}{\longrightarrow}\,\ 
{\langle   P_q \vert \widetilde O_{i}(\mu)  \vert   P_q \rangle \over
 b_i \ \vert \langle 0 \vert \widetilde A_0 \vert   P_q \rangle \vert^2} \ \equiv \widetilde B_i(\mu) \ ,
\eea
where $b_i \in \{ 8/3, -5/3, 2,2/3 \}$ for $i\in \{1,2,4,5\}$. In this case, 
the fixed time has been chosen to be $t_2=35$. In fig.~\ref{figH}, we show 
the quality of the signals for all four $\widetilde R_i(t_1)$. 
On the plateaus we fit to a constant and thus obtain the values of 
the corresponding parameters $\widetilde B_i(\mu)$. Our data is 
renormalized in the $\msbar$(NDR) scheme, after matching
the lattice regularization scheme onto the $\msbar$(NDR) by using 
one-loop boosted perturbation theory, as explained in great 
detail in ref.~\cite{GR1}. The matching scale $\mu = q^\ast$ is varied 
between $2/a \leq q^\ast \leq \pi/a$, and the results are  
run to $\mu = m_b = 4.6$~GeV. The spread of values is assigned to the 
systematic uncertainty. As in the previous subsection, 
all 4 operators are linearly extrapolated (interpolated) in the the light quark mass
to up/down (strange). The resulting values are 
listed in table~\ref{tab3}. 
\begin{table}[h!] 
\begin{center} 
\begin{tabular}{|c|c|c|} \hline 
{\sl \small Light quark}&
{\sl \small q=up/down}& 
{\sl \small q=strange} \\
\hline 
\hline 
{\phantom{\Huge{l}}}\raisebox{-.2cm}{\phantom{\Huge{j}}}  
\hspace*{-5mm} $\widetilde B_1 (m_b)$ & 
0.89(4) & 0.89(4)  \\ 
{\phantom{\Huge{l}}}\raisebox{-.2cm}{\phantom{\Huge{j}}}
\hspace*{-5mm} $\widetilde B_2 (m_b)$ & 
 0.82(4) & 0.83(3) \\ 
{\phantom{\Huge{l}}}\raisebox{-.2cm}{\phantom{\Huge{j}}}
\hspace*{-5mm} $\widetilde B_4 (m_b)$ & 
1.07(4) & 1.07(4)  \\
{\phantom{\Huge{l}}}\raisebox{-.2cm}{\phantom{\Huge{j}}}
\hspace*{-5mm} $\widetilde B_5 (m_b)$ & 
2.37(10) & 2.40(10)  \\ \hline
\end{tabular} 
\caption{\label{tab3}
\small{\sl $\tilde B$-parameters, extracted from our HQET data by using the 
boosted perturbative 1-loop matching onto the continuum 
$\msbar$(NDR) renormalization scheme. }}
\end{center}
\vspace*{-.3cm}
\end{table}
In that table errors are statistical only, obtained by using the standard
jackknife procedure.

\section{Extrapolation to the $B$-mesons \label{sec3}}

Armed with ``raw'' results obtained in full QCD (table~\ref{tab2}) and in HQET 
 (table~\ref{tab3}), we now discuss the 
extrapolation of the QCD results to the physical $B$-meson mass. 
The aim of this section is to provide a consistent way to constrain that extrapolation 
by the static HQET results in order to reduce the systematic 
uncertainties.

A common wisdom is to follow the HQET scaling laws, 
according to which every $B$-parameter scales
with the inverse heavy quark (meson) mass as a constant, and to  
extrapolate to the desired heavy meson mass. The (unknown) 
$1/m_P$ corrections are to be determined from the fit with our data.
To use these scaling laws, however, 
one first need to relate the matrix elements of the QCD operators, 
$\langle O_i (\mu)\rangle$ from eq.~(\ref{basisc}), to the HQET 
ones, $\langle \widetilde O_i(\mu)\rangle$ of eq.~(\ref{basish})~\cite{Eichten:1990zv}.
This matching is made in perturbation theory at some suitably chosen 
renormalization scale, for example $\mu = m_b$.
The matching is crucial since the 
anomalous dimensions for these operators in the 
two theories (QCD and HQET) differ. 
Moreover, when dealing with the 4-fermion operators, it is 
important that the matching between the two theories is made at NLO accuracy 
because it is at this order that the scheme
can be fully specified (leading order anomalous dimensions are universal). 
Before entering the details of that matching, we now outline the basic 
strategy that must be followed.

Matching of the QCD operators, renormalized at some high scale $\mu \gg m_P$, 
and the HQET ones, renormalized at some low scale $\mu^\prime \ll m_P$, is made
at $\mu = m_{P}$ by using the following expression~\footnote{
Instead of matching at the point corresponding to the mass of the heavy quark, 
we choose to do it at the mass of the heavy-light meson, $m_P$.}
\bea \label{match1}
{\bf W}_{QCD}^T[m_P, \mu]^{-1}\ 
\langle \vec O (\mu) \rangle_{m_P} \ =\  C (m_{P})\ {\bf W}_{HQET}^T[m_P, \mu^\prime ]^{-1} 
\ \langle \vec{\widetilde O} (\mu^\prime )\rangle + {\cal
O}\left({1\over m_{P}}\right)
\ +\ \dots  
\eea
where ${\bf W}^T_{QCD}[\mu_2, \mu_1]^{-1}$ is the matrix encoding 
the full QCD evolution from a scale $\mu_1$ to $\mu_2$ of all five $\Delta B=2$ 
operators which are, for convenience, collected in a five-component vector 
$\langle \vec O (\mu) \rangle_{m_P}$. 
Likewise for ${\bf W}^T_{HQET}[\mu_2, \mu_1]^{-1}$ in HQET. These matrices 
will be specified later on. 
We will be working in the $\msbar$(NDR) scheme in which the matrices of the anomalous dimension
coefficients are known at NLO in both theories.
Hence, the matching matrix, $C(m_{P})= 1 + 
\sum_n  c^{(n)} [\alpha_s(m_{P})/4\pi]^n$, is also known at NLO, 
{\it i.e.}  $c^{(1)}$ is completely determined~\cite{GR2}.

On the HQET side, we also consider five operators,  
$\langle \vec {\widetilde O} (\mu)\rangle$, where we add the 
operator $\widetilde O_3$ by means of eq.~(\ref{eq3}). 
In this way the matching matrix $c^{(1)}$ is squared ($5\times 5$). 
We now put all the evolution expressions appearing in eq.~(\ref{match1}) on its l.h.s.
\bea \label{match2}
\hspace*{-8mm}{\bf W}_{HQET}^T[\mu^\prime ,m_P]^{-1} C^{-1} (m_{P}){\bf W}^T_{QCD}[m_P, \mu]^{-1} 
\langle \vec O (\mu) \rangle_{m_P} &=& \cr
&&\cr
\hspace*{-8mm}\biggl( \ \underbrace{  {\bf W}^T_{QCD}[m_P, \mu] C (m_{P}) {\bf
W}^T_{HQET}[m_P, \mu^\prime]^{-1} \ }_{\displaystyle{{\cal M}_4[m_P, \mu, \mu^\prime]}}\biggr)^{-1} 
\langle \vec O (\mu) \rangle_{m_P} &=& \langle \vec{\widetilde O} (\mu^\prime)\rangle + {\cal
O}\left({1\over m_{P}}\right)
+ \dots  
\eea
so that the l.h.s. manifestly satisfies the HQET scaling laws which are the 
intrinsic property of the r.h.s. One proceeds similarly for the bilinear
operators to define the matching constants ${\cal M}_2[m_P, \mu, \mu^\prime ]$. In terms of 
$B$-parameters, eq.~(\ref{match2}) then reads
\bea
{\cal M}^{2}_2[m_P, \mu, \mu^\prime ] \  \biggl( {\bf b}^{-1} {\cal M}^{-1}_4[m_P, \mu, \mu^\prime] {\bf
b}\biggr) \ \vec B (\mu) = 
\vec {\widetilde B} (\mu^\prime )  +  {\cal O}\left({1\over m_{P}}\right) + \dots  
\eea
where ${\bf b}$ is the diagonal matrix of the coefficients appearing in the
definitions~(\ref{params}), {\it i.e.} ${\bf b} = {\rm diag}( 8/3, -5/3, 1/3, 
2, 2/3)$, and by $\vec B (\mu)$ ($\vec {\widetilde B} (\mu^\prime)$) 
we designated the vector column of our five $B$-parameters. 

Based on the above discussion, a simple recipe can be applied to our data, namely
evolve to the same $\mu=\mu^\prime$ and create the quantity 
\bea \label{defPHI}
\vec \Phi (m_P,\mu) ={\cal M}^{2}_2[m_P, \mu] \, \biggl( {\bf b}^{-1} {\cal M}^{-1}_4[m_P, \mu] 
{\bf b}\biggr) \cdot  \vec B (\mu)
\eea
which can be fit either freely as
\bea \label{fit10}
\vec \Phi (m_P,\mu) = \vec a_0(\mu) +  {\vec a_1(\mu) \over m_P}\,,
\eea
where  $\vec a_0(\mu)$ and $\vec a_1(\mu)$ are the fit parameters,  
or by constraining it by the static HQET results, {\it i.e.}
\bea \label{fit20}
\vec \Phi (m_P,\mu) = \vec a_0^\prime(\mu) + {\vec a_1^\prime(\mu) \over m_P}+ 
{\vec a_2^\prime(\mu) \over m_P^2}\;,
\eea
where the coefficient $\vec a_0^\prime(\mu)$ is constrained by the static value,
 $\vec {\widetilde B}(\mu)$, so that one can probe the term 
${\cal O}(1/m_P^2)$. 
As a result of these two procedures, we obtain the HQET values of
the $B$-parameters, {\it i.e.} $\vec \Phi (m_{B_{s/d}},\mu)$, which are 
then to be matched back onto their QCD counterparts.

To keep the expressions as short as possible, we will now split the 
discussion into two pieces: we will first discuss the extrapolation of the first
three $B$-parameters and then the last two. This can be done because 
all the matrices, ${\bf W}^T_{QCD}[\mu_2, \mu_1]$, ${\bf W}^T_{HQET}[\mu_2, \mu_1]$ and $C (\mu)$ 
are the block-matrices of the form $[3\times 3] \oplus [2\times 2]$.

\subsection{ Getting the physical results for $B_{1,2,3}^\msbar(m_b)$}

At the leading order in perturbation theory, the anomalous dimensions 
for our $B$-parameters in the RI/MOM and $\msbar$ schemes
are the same. This is not the case at NLO, and in order to proceed we need to
convert our RI/MOM results from table~\ref{tab2} into the $\msbar$(NDR) scheme. 
It is crucial to specify the set of evanescent operators or the Dirac 
projectors used to renormalize the operators because only with this information 
at hand, the $\msbar$(NDR) scheme is unambiguously defined~\cite{nierste}. 
In this subsection, we will use the $\msbar$(NDR) scheme of ref.~\cite{beneke} (see
eqs.~(13-15) of their paper) in which the Wilson coefficients for the 
SM expression for the $(\Delta \Gamma/\Gamma)_{B_s}$ have been calculated at 
NLO. Therefore, the results for the $B$-parameters that will be
presented in this subsection can be directly combined with the Wilson
coefficients of ref.~\cite{beneke}.

The conversion of the operators $O_{1,2,3}^\ri (\mu)$ and $P_5^\ri(\mu)$ 
to the $\msbar$ scheme is provided by the following expressions
\bea
&&
\left(
\begin{array}{c}
\langle O_1(\mu)\rangle  \\
\langle O_2(\mu)\rangle  \\
\langle O_3(\mu)\rangle  \\
\end{array}
\right)^\msbar =  \left[\; \mathbb{I} \ +\ r_{123}  {\alpha_s(\mu)\over 4 \pi} \; \right] \ 
\left(
\begin{array}{c}
\langle O_1(\mu)\rangle  \\
\langle O_2(\mu)\rangle  \\
\langle O_3(\mu)\rangle  \\
\end{array}
\right)^\ri \;,\cr
&&\cr
&&\cr
&&\langle P_5(\mu)\rangle^\msbar = \biggl( 1 + r_P {\alpha_s(\mu)\over 4 \pi} \biggr) \ 
\langle P_5 (\mu)\rangle^\ri \;,
\eea
where the NLO matching coefficients are given by~\cite{ciuchini,buras,GR2}
\bea
\hspace*{10mm}r_P={16\over 3}\;, \hspace*{10mm} 
r_{123} ={1\over 9}\ \left( \begin{array}{ccc}
-{42}\ +\ 72 \log 2 & 0 & 0 \\
0 &  61 + 44 \log 2 & -7 + 28 \log 2  \\
0 & -25 + 28 \log 2 & -29 + 44 \log 2   \\
\end{array} \right)\;.
\eea
To get the central values, we will convert the results from table~\ref{tab2}, 
obtained at $\mu = 2.8(1)$~GeV and run them to $\mu = m_b = 4.6$~GeV~\cite{mb}. 
It should be noted that the matching $\ri \to \msbar$(NDR), made in 
ref.~\cite{ape2}, was incorrect because the $\msbar$ scheme was not the one 
of ref.~\cite{beneke}, but rather the one of ref.~\cite{buras}. 
Although the numerical differences are very small, the physical
results presented in ref.~\cite{ape2} are not fully consistent because 
the matrix elements
matched onto the $\msbar$ scheme of ref.~\cite{buras} were combined with the
Wilson coefficients of ref.~\cite{beneke}. The correct physical results 
were presented in ref.~\cite{budapest}.

The evolution from the scale $\mu$ to $m_b$, in this $\msbar$ scheme, is
described by~\cite{beneke}  
\bea \label{evol123}
&&\left(
\begin{array}{c}
O_1(m_b) \\
O_2(m_b) \\
O_3(m_b) \\
\end{array}
\right)^{\msbar}
= {\bf W}_{QCD}^T[m_b,\mu]^{-1} \left(
\begin{array}{c}
O_1(\mu) \\
O_2(\mu) \\
O_3(\mu) \\
\end{array}
\right)^{\msbar} \eea
where the operator ${\bf W}_{QCD}[m_b,\mu] = M(m_b) U(m_b,\mu) M^{-1}(\mu)$ 
contains the information on the evolution obtained at the leading ($U(\mu,m_b)$) and 
the NLO ($M(\mu)$) in perturbation theory. For our purpose, it is 
convenient to write the evolution matrix in the following form
\bea
\label{scale5}
&&{\bf W}_{QCD}[m_b,\mu ] = w(m_b) w^{-1}(\mu)\,,  \eea
where 
\bea 
&&  w(\mu)\ =\ M(\mu) \ \alpha_s(\mu)^{-\gamma_0^T/2\beta_0}\ , 
\eea
and $\beta_0 = 11 - 2 n_F/3$. 
The scheme independent, one-loop anomalous dimension matrix is 
\[
{\gamma_0} =
\left(
\begin{array}{rrr}
4 &0&0\\
0&  -28/3& 4/3 
\\
0 & 16/3 & 32/3 
\\
\end{array}
\right)  \;,
\]
whereas the NLO contribution
\bea
M(\mu)\ =\ \mathbb{I} \  +\ {J}_{123}^{\msbar}\ {\alpha_s(\mu)\over 4 \pi}\; 
\eea
is encoded in the  matrix 
\bea
&&{J}_{123}^{\msbar}\ = 
 \ \left( \begin{array}{ccc}  
{\phantom{\Huge{l}}}\raisebox{-.4cm}{\phantom{\Huge{j}}}
 {\displaystyle{ 485}\over \displaystyle{242}} &\;0\; & \; 0  
\\
{\phantom{\Huge{l}}}\raisebox{-.4cm}{\phantom{\Huge{j}}}
\; 0\; & -{\displaystyle{4592}\over \displaystyle{9075}}  &\;  
-{\displaystyle{19083}\over \displaystyle{3025}} \\
{\phantom{\Huge{l}}}\raisebox{-.4cm}{\phantom{\Huge{j}}}
\; 0\; & {\displaystyle{1612}\over \displaystyle{3025}}  &\;  
-{\displaystyle{36233}\over \displaystyle{9075}} 
\end{array} \right) \; .  
\eea
We have set $n_F=0$, since our lattice results are obtained in 
the quenched approximation.

The evolution of the pseudoscalar density is given by
\bea
\langle P_5(m_b)\rangle^\msbar = \left(\alpha_s(\mu)\over \alpha_s(m_b)\right)^{-\gamma_P/2\beta_0}
\left[\  1 + {\alpha_s(\mu) - \alpha_s(m_b)\over 4 \pi} J_P^{\msbar} \right]\ \langle P_5 
(\mu)\rangle^\msbar \;,
\eea
where $\gamma_P =-8$ and $J_P^{\msbar} = 998/363$, for $n_F=0$.

With all of the above formulae at hand, we convert our 
$B$-parameters from table.~\ref{tab2} to the $\msbar$ scheme (at $\mu = 2.8(1)$~GeV), run them 
to $\mu = m_b$, and list their values in table~\ref{tab4}.
\begin{table}[h!] 
\begin{center} 
\begin{tabular}{|c|c|c|c||c|c|c|} \cline{2-7}
\multicolumn{1}{l|}{}&\multicolumn{3}{c||}
{\sl \small q=up/down}& 
\multicolumn{3}{c|}{\sl \small q=strange} \\
\hline 
{\phantom{\Huge{l}}}\raisebox{-.2cm}{\phantom{\Huge{j}}}  
\hspace*{-5mm} $\kappa_Q$ & 
0.125 & 0.122 & 0.119 & 0.125 & 0.122 & 0.119\\ \hline 
{\phantom{\Huge{l}}}\raisebox{-.2cm}{\phantom{\Huge{j}}}  
\hspace*{-5mm} $m_{P} \ [\gev ]$ & 
1.75(9) & 2.02(10)& 2.26(12) & 1.85(8) & 2.12(9)& 2.36(10)\\ \hline 
{\phantom{\Huge{l}}}\raisebox{-.2cm}{\phantom{\Huge{j}}}  
\hspace*{-5mm} $B_1^\msbar(m_b)$ & 
0.891(21) & 0.913(26)  & 0.920(27) & 
0.900(16) & 0.916(18)  & 0.919(16)\\ 
{\phantom{\Huge{l}}}\raisebox{-.2cm}{\phantom{\Huge{j}}}
\hspace*{-5mm} $B_2^\msbar (m_b)$ & 
 0.745(29) & 0.771(29)  & 0.824(26)& 
0.788(21) & 0.817(21)  & 0.833(18)\\ 
{\phantom{\Huge{l}}}\raisebox{-.2cm}{\phantom{\Huge{j}}}
\hspace*{-5mm} $B_3^\msbar (m_b)$ & 
0.896(41) & 0.889(43) & 0.943(41)& 
0.902(27) & 0.918(26)  & 0.921(26) \\
 \hline
\end{tabular} 
\caption{\label{tab4}
\small{\sl $B$-parameters, extracted from our lattice QCD data after conversion to the 
$\msbar$(NDR) scheme of ref.~\cite{beneke} at $\mu = 2.8(1) \gev$ and running to $\mu = m_b$.}}
\end{center}
\vspace*{-.3cm}
\end{table}

The next step is to match the $B$-parameters from table~\ref{tab4} from QCD to the 
HQET where we can use the heavy quark scaling laws to
extrapolate each $B$-parameter to the physical point, {\it i.e.} to $m_{B_{d/s}}$. 
The crucial ingredient that enters the matrix ${\cal M}^{-1}_4[ m_P,m_b]\equiv 
{\cal M}^{-1}_4[ m_P,m_b,m_b]$ of
eq.~(\ref{match2}), and hence the quantity 
$\vec \Phi(m_P,m_b)$ in eq.~(\ref{defPHI}), is the matching matrix 
 $C(m_P)$ which relates the QCD operators, 
computed in the $\msbar$(NDR)
scheme of ref.~\cite{beneke}, onto the HQET ones computed in the $\msbar$(NDR)
scheme of ref.~\cite{GR1} 
(and vice versa). In ref.~\cite{GR2} it has
been shown that (at NLO) this matrix has the following form:
\bea
C_{123} (m_P) = \mathbb{I}   + {c}^{(1)}_{123}\ {\alpha_s(m_P)\over 4 \pi}\; ;\hspace*{16mm}
{c}^{(1)}_{123}=\left(
\begin{array}{ccc}
-14 & -8 & 0 \\
0 & 61/12 & -13/4 \\
0 & -77/12 & -121/12 \\
\end{array}
\right)\;,
\eea
where we introduced the index $``123"$ since we consider only these operators here.

The last piece of information  needed to 
construct ${\cal M}^{-1}_4[ m_P,m_b]$ of
eq.~(\ref{match2}), is the evolution 
operator in the HQET at NLO. In a notation analogous to eq.~(\ref{scale5}) we
have 
\bea
&& \langle \vec{\widetilde O} (m_b) \rangle = {\bf W}_{HQET}^T[m_b,\mu]^{-1} \langle \vec{\widetilde
O} (\mu) \rangle \cr
&& \cr
&& \cr
&&  {\bf W}_{HQET}[m_b,\mu] = \widetilde w(m_b) \widetilde w^{-1}(\mu)\,,  \eea
where 
\bea 
&&  \widetilde w(\mu)\ =\ \widetilde M(\mu) \ \alpha_s(\mu)^{-{\widetilde \gamma_0}^T/2\beta_0}\ .
\eea
In this case~\cite{GR2} 
\[
{\widetilde \gamma_0} =
\ -{8 \over 3}\ 
\left(
\begin{array}{rrr}
3 & 0 & 0 \\
0 & 2 & 1 \\
0 & 1 & 2 \\
\end{array}
\right)  \;,
\]
\bea
\widetilde M(\mu)\ =\ \mathbb{I} \  +\ {\widetilde J}_{123}^{\ \msbar}\
{\alpha_s(\mu)\over 4 \pi}\; ,
\eea
\bea
\widetilde J_{123}^{\ \msbar} = 
\left(
\begin{array}{ccc}
{\phantom{\Huge{l}}}\raisebox{-.4cm}{\phantom{\Huge{j}}}
\displaystyle{ -{317\over 1089} + {68 \pi^2 \over 297}} & 
\displaystyle{ -{2639\over 8712} + {\pi^2 \over 99}} &  0 \\
{\phantom{\Huge{l}}}\raisebox{-.4cm}{\phantom{\Huge{j}}}
0 & \displaystyle{ -{3181\over 4356} + {74 \pi^2 \over 297}} & 
\displaystyle{ {1913\over 4356} - {2 \pi^2 \over 99}} \\
{\phantom{\Huge{l}}}\raisebox{-.4cm}{\phantom{\Huge{j}}}
0 & \displaystyle{ {2639\over 4356} - {2 \pi^2 \over 99}} & 
\displaystyle{ -{3907\over 4356} + {74 \pi^2 \over 297}} \\
\end{array}
\right) \;.
\eea
where, as before, we have set $n_F=0$.

To obtain the quantity $\vec \Phi(m_P,m_b)$ of eq.~(\ref{defPHI}), one also
needs ${\cal M}^{-1}_2[ m_P,m_b]$. At NLO this information can be 
extracted from ref.~\cite{broadhurst}. The matching of the axial current and of the pseudoscalar
density is given by
\bea
&&A_0 \ = \ \left[ \ 1\ -\ {8 \over 3} {\alpha_s(\mu)\over 4
\pi}\ \right] \ \widetilde A_0(\mu)\;, \cr
&&\cr
&&P_5(\mu)\ =\ \left[ \ 1\ +\ {8 \over 3} {\alpha_s(\mu)\over 4
\pi}\right] \ \widetilde A_0(\mu)  \;.
\eea
while the expression for the running of the axial current (for 
$n_F=0$) is 
\bea
&&\widetilde A_0(m_b) = \left( {\alpha_s(\mu)\over \alpha_s(m_b)}\right)^{-\widetilde \gamma_A/2\beta_0} \ 
\left[ 1 - \left( 
{ {439\over 1089} - {28 \pi^2 \over 297}  }\right){\alpha_s(\mu) - \alpha_s(m_b)\over 4
\pi}\right]  \widetilde A_0(\mu) \;,
\eea
where the leading order anomalous dimension is given by 
$\widetilde \gamma_A = -4$.

\begin{figure}[h!]
\begin{center}
\begin{tabular}{c c c}
\hspace*{-11mm}&\includegraphics[height=14cm,width=10cm]{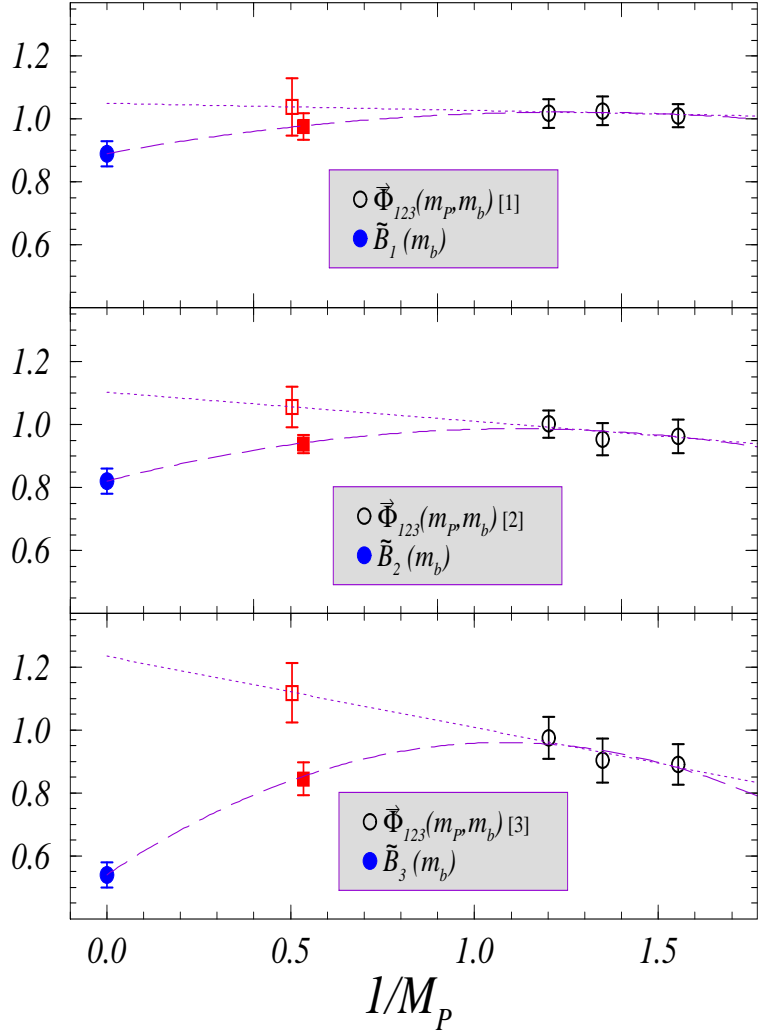} &  \\
\end{tabular}

\caption{\label{fig33}{\sl \small Extrapolation to the physical 
$B_d$ meson mass (squared symbols) in the inverse heavy meson mass. 
The unconstrained linear extrapolation for each of the components
of the vector $\vec \Phi_{1,2,3}(m_{P},m_b)$ from our data (empty circles) to $\vec
\Phi_{1,2,3}(m_{B_d},m_b)$ (empty square) is depicted by the dotted line. The result of the constrained
extrapolation (filled squares) by the static HQET bag-parameters (filled circles) is marked by the 
dashed line. $i^{th}$ component of the vector is marked by $[i]$.}}
\end{center}
\end{figure}

By combining all of the above ingredients, we form 
the quantities  $\vec \Phi_{1,2,3}(m_P, m_b)$, use  
eq.~(\ref{fit10}), and extrapolate to $m_{B_{d/s}}$. 
The result of that extrapolation is:
\bea
\vec \Phi_{1,2,3}(m_{B_d},m_b) = 
\left(
\begin{array}{c}
 1.038(91) \\
 1.055(64) \\
 1.117(94) \\
\end{array}
\right)  \;,\hspace*{12mm} 
\vec \Phi_{1,2,3}(m_{B_s},m_b) = 
\left(
\begin{array}{c}
 1.009(36) \\
 1.045(24) \\
 1.070(48) \\
\end{array}
\right)  \;,
\eea
where we wrote separately the results of the extrapolation of our data with the light quark
extrapolated to $d$ (left), and those with the light quark interpolated to $s$ (right).  
By including the $\widetilde B$-parameters from table~\ref{tab3}, we 
fit our data to eq.~(\ref{fit20}), from which we get
\bea \label{fittos}
\vec \Phi_{1,2,3}(m_{B_d},m_b) = 
\left(
\begin{array}{c}
 0.976(42) \\
 0.943(28) \\
 0.845(52) \\
\end{array}
\right)  \;,\hspace*{12mm} 
\vec \Phi_{1,2,3}(m_{B_s},m_b) = 
\left(
\begin{array}{c}
 0.967(21) \\
 0.955(17) \\
 0.857(31) \\
\end{array}
\right)  \;.
\eea
These two extrapolations, for each component of the vector 
$\vec \Phi_{1,2,3}(m_{P},m_b)$, are illustrated in fig.~\ref{fig33} for 
the case of the light $d$-quark. Results of that extrapolation are 
the $B$-parameters in HQET, which we then match back onto their QCD values to get our final results for
$B$-parameters. These numbers are presented in table~\ref{tab123}. 

An important issue to be mentioned is the treatment of the statistical
errors when constraining by the static HQET results because the three points obtained 
in QCD are correlated among themselves and are uncorrelated from the one obtained 
in HQET. 
Although it may 
look trivial, we prefer to mention how these errors have been 
treated in this work.~\footnote{For a clear discussion about the treatment of
the statistical errors in such a situation, see ref.~\cite{lepage}.}
For each jack of our QCD data  
we form the so-called augmented $\chi^2_a$ by including the HQET value 
$\tilde B_j$ and its error $\tilde \sigma_j$ as:
\bea
(\chi^2_a)_j \ =\  \sum_i \left[ { \, \Phi(m_{P_i})[j] - (a_0^\prime)_j  - 
{ \displaystyle{(a_1^\prime)_j}\over \displaystyle{m_{P_i}} } - 
{\displaystyle{(a_2^\prime)_j}\, \over \displaystyle{m_{P_i}^2}} 
\over \sigma_j[\Phi (m_{P_i})] } \right]^2 
+  \biggl( {\tilde B_j - (a_0^\prime )_j \over \tilde \sigma_j}  \biggr)^2\;,
\eea
where ``$[j]$" denotes the $j^{th}$ component of the vector $\vec
\Phi(m_{P_i})$, the error of which is
$\sigma_j[\Phi (m_{P_i})]$ for the 
$i^{th}$ value of our three heavy-light meson masses, $m_{P_i}$. 
By minimizing $(\chi^2_a)_j$, we find the values of the parameters 
$(a_{0,1,2}^\prime)_j$, where $(a_{0}^\prime)_j$ is constrained 
by the prior knowledge of the static result. In this way we get
the value for $\Phi(m_{B_{d/s}})[j]$, for each jack of our QCD data. 
The final error, $\sigma_j(\Phi (m_{B_{d/s}}))$, that we quoted in 
eq.~(\ref{fittos}), is obtained as a standard error over all jacks ($N_{JK}$)
\bea
\sigma_j(\Phi (m_{B_{d/s}})) = \sqrt{ { N_{JK}-1 \over N_{JK} }\ \left[
\sum_{k=1}^{N_{JK}} \left( \Phi_k(m_{B_{d/s}})[j]\right)^2 - 
{ 1 \over N_{JK} }\left(\sum_{k=1}^{N_{JK}} \Phi_k(m_{B_{d/s}})[j]\right)^2 \right] 
}\;.
\eea
\begin{table}[h!] 
\begin{center} 
\begin{tabular}{|c|c|c||c|c|} \cline{2-5}
\multicolumn{1}{l|}{}&\multicolumn{2}{c||}
{\sl \small unconstrained}& 
\multicolumn{2}{c|}{\sl \small constrained} \\
\hline 
{\phantom{\Huge{l}}}\raisebox{-.2cm}{\phantom{\Huge{j}}}  
\hspace*{-5mm} {\sl \small Light quark} & 
$d$ & $s$  & $d$ & $s$ \\ \hline  
{\phantom{\Huge{l}}}\raisebox{-.2cm}{\phantom{\Huge{j}}}  
\hspace*{-5mm} $B_1^\msbar(m_b)$ & 
0.938(81) & 0.905(32) & 0.875(37)  & 0.867(18)\\ 
{\phantom{\Huge{l}}}\raisebox{-.2cm}{\phantom{\Huge{j}}}
\hspace*{-5mm} $B_2^\msbar (m_b)$  & 
0.923(32)& 0.915(30) & 0.826(25)  & 0.836(15)\\ 
{\phantom{\Huge{l}}}\raisebox{-.2cm}{\phantom{\Huge{j}}}
\hspace*{-5mm} $B_3^\msbar (m_b)$ & 
1.192(101)& 1.141(52) & 0.901(56)  & 0.914(33) \\
 \hline
\end{tabular} 
\caption{\label{tab123}
\small{\sl Final results for the first three $B$-parameters defined in
eq.~(\ref{params}), 
in the $\msbar$(NDR) scheme of ref.~\cite{beneke} at $\mu = m_b$. }}
\end{center}
\vspace*{-.3cm}
\end{table}

From table~\ref{tab123} we see that by extrapolating the $B$-parameters 
from the range of masses accessed from our lattice 
to the $m_{B_{d/s}}$, without including the static HQET results, 
we always overshoot the ones that are obtained by including the static 
HQET values. This is especially pronounced for the parameter $B_3(m_b)$.
At this point it is not clear whether this is the real physical effect, 
or it is due to the lattice artefacts: {\it e.g.} our heavier mesons may be more 
subject to ${\cal O}(a)$ effects, our HQET data are only perturbatively renormalized etc. 
Further research  is needed to clarify this issue.

\subsection{Physical results for $B_{4,5}^\msbar(m_b)$}

\begin{figure}[h!]
\begin{center}
\begin{tabular}{c c c}
\hspace*{-11mm}&\includegraphics[height=9.4cm,width=10cm]{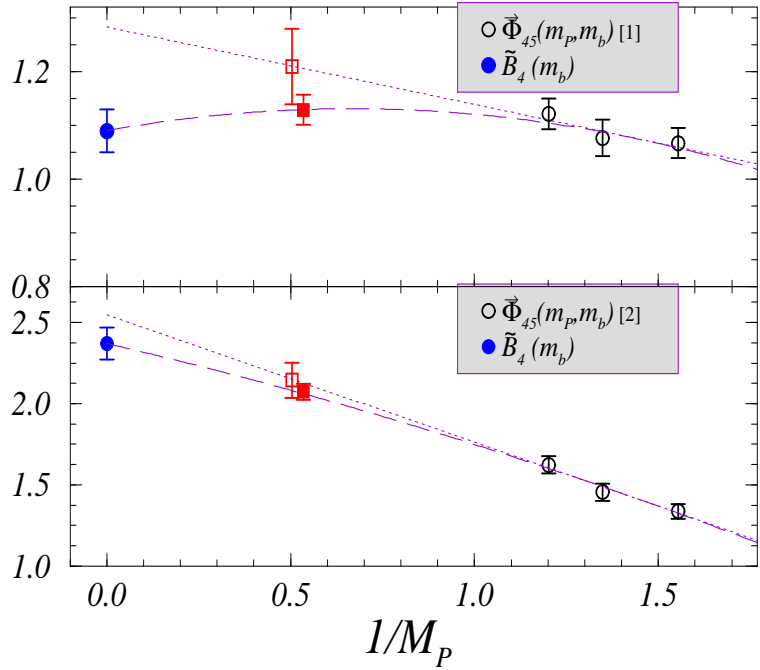} &  \\
\end{tabular}

\caption{\label{fig44}{\sl \small The same as fig.~\ref{fig33}
but for $\vec \Phi_{4,5}(m_{P},m_b)$.}}
\end{center}
\end{figure}

As in the previous section, we first convert our directly computed 
$B^\ri_{4,5}\to B^\msbar_{4,5}$. We choose the $\msbar$(NDR) scheme 
of ref.~\cite{buras}, according to which 
\bea
&&
\left(
\begin{array}{c}
\langle O_4(\mu)\rangle  \\
\langle O_5(\mu)\rangle  \\
\end{array}
\right)^\msbar =  \left[\; \mathbb{I} \ +\ r_{45}  {\alpha_s(\mu)\over 4 \pi} \; \right] \ 
\left(
\begin{array}{c}
\langle O_4(\mu)\rangle  \\
\langle O_5(\mu)\rangle  \\
\end{array}
\right)^\ri \;,
\eea
with the NLO matching coefficient given by~\cite{buras}
\bea 
r_{45} =-{2\over 3}\ \left( \begin{array}{cc}
  -17 + \log 2 & 3 (1 - \log 2)  \\
 -3 (1 + \log 2) &  1 + \log 2    \\
\end{array} \right)\;.
\eea
As in the previous subsection, we convert the results from table~\ref{tab2} 
at $\mu = 2.8(1)$~GeV and evolve them in the $\msbar$ scheme to the scale 
$\mu = m_b = 4.6$~GeV. 
The evolution is described by an equation analogous to eq.~(\ref{evol123})
in which the one-loop anomalous dimension now reads~\cite{buras}
\[
{\gamma_0} =
\left(
\begin{array}{rr}
-16 &0\\
 -6& 2\\
\end{array}
\right)  \;,
\]
whereas the NLO part ($n_F=0$) is 
\bea
M(\mu)\ =\ \mathbb{I} \  +\ {J}_{45}^{\msbar}\ {\alpha_s(\mu)\over 4 \pi}\;, 
\hspace*{11mm} 
{J}_{45}^{\msbar}\ = 
 \ \left( \begin{array}{cc}  
{\phantom{\Huge{l}}}\raisebox{-.4cm}{\phantom{\Huge{j}}}
 {\displaystyle{ 24379}\over \displaystyle{ 5808}}  &   
{\displaystyle{ 5895}\over \displaystyle{ 1936}} \\
{\phantom{\Huge{l}}}\raisebox{-.4cm}{\phantom{\Huge{j}}}
 {\displaystyle{45}\over \displaystyle{16}}  &   
-{\displaystyle{5807}\over \displaystyle{5808}} 
\end{array} \right) \; .  
\eea
Our $B_{4,5}$-parameters, in the $\msbar$ scheme and at $\mu = m_b$, are 
given in table~\ref{tab5}.
\begin{table}[h!] 
\begin{center} 
\begin{tabular}{|c|c|c|c||c|c|c|} \cline{2-7}
\multicolumn{1}{l|}{}&\multicolumn{3}{c||}
{\sl \small q=up/down}& 
\multicolumn{3}{c|}{\sl \small q=strange} \\
\hline 
{\phantom{\Huge{l}}}\raisebox{-.2cm}{\phantom{\Huge{j}}}  
\hspace*{-5mm} $\kappa_Q$ & 
0.125 & 0.122 & 0.119 & 0.125 & 0.122 & 0.119\\ \hline 
{\phantom{\Huge{l}}}\raisebox{-.2cm}{\phantom{\Huge{j}}}  
\hspace*{-5mm} $m_{P} \ [\gev ]$ & 
1.75(9) & 2.02(10)& 2.26(12) & 1.85(8) & 2.12(9)& 2.36(10)\\ \hline 
{\phantom{\Huge{l}}}\raisebox{-.2cm}{\phantom{\Huge{j}}}  
\hspace*{-5mm} $B_4^\msbar(m_b)$ & 
1.098(22) & 1.074(25)  & 1.126(21) & 
1.119(16) & 1.134(17)  & 1.131(13)\\ 
{\phantom{\Huge{l}}}\raisebox{-.2cm}{\phantom{\Huge{j}}}
\hspace*{-5mm} $B_5^\msbar (m_b)$ & 
 1.235(25) & 1.308(31)  & 1.442(30)& 
1.288(20) & 1.390(24)  & 1.466(21)\\ 
 \hline
\end{tabular} 
\caption{\label{tab5}
\small{\sl $B$-parameters, extracted from our lattice QCD data and converted to the 
$\msbar$(NDR) scheme of ref.~\cite{buras}.}}
\end{center}
\vspace*{-.3cm}
\end{table}
The matching onto the corresponding operators in HQET 
is made through~\cite{GR2} 
\bea
C_{45} (m_P) = \mathbb{I}   + 
{c}^{(1)}_{45}\ {\alpha_s(m_P)\over 4 \pi}\; ,\hspace*{16mm}
{c}^{(1)}_{45}= {1\over 2} \left(
\begin{array}{cc}
17 & -11  \\
7 & -21 \\
\end{array}
\right)\;.
\eea
As for the evolution of these operators in HQET, the matrix of the 
leading order anomalous dimension coefficients is  
\[
{\widetilde \gamma_0} =
\ - \ 
\left(
\begin{array}{rr}
7 & 3 \\
3 & 7 \\
\end{array}
\right)  \;,
\]
while the NLO contribution ($n_F=0$) reads~\cite{GR2}
\bea
\widetilde M(\mu)\ =\ \mathbb{I} \  +\ {\widetilde J}_{45}^{\msbar}\ {\alpha_s(\mu)\over 4 \pi}\; ,
\hspace*{11mm} 
\widetilde J_{45}^{\msbar} = 
\left(
\begin{array}{cc}
{\phantom{\Huge{l}}}\raisebox{-.4cm}{\phantom{\Huge{j}}}
\displaystyle{ -{833\over 726} + {74 \pi^2 \over 297}} & 
\displaystyle{ {1987\over 2178} - {2 \pi^2 \over 99}}  \\
{\phantom{\Huge{l}}}\raisebox{-.4cm}{\phantom{\Huge{j}}}
 \displaystyle{{1987\over 2178} - {2 \pi^2 \over 99}} & 
\displaystyle{ -{833\over 726} + {74 \pi^2 \over 297}} \\
\end{array}
\right) \;.
\eea
We use all the above formulae to create the quantities 
$\vec \Phi_{4,5}(m_P,m_b)$~(\ref{defPHI}) and extrapolate in the heavy 
meson mass by using eq.~(\ref{fit10}). The results are
\bea
\vec \Phi_{4,5}(m_{B_d},m_b) = 
\left(
\begin{array}{c}
 1.189(70) \\
 2.144(108) \\
\end{array}
\right)  \;,\hspace*{12mm} 
\vec \Phi_{4,5}(m_{B_s},m_b) = 
\left(
\begin{array}{c}
 1.184(25) \\
 2.158(60) \\
\end{array}
\right)  \;.
\eea
After incorporating the values for $\widetilde B_{4,5}(m_b)$ given in 
table~\ref{tab3}, the fit to eq.~(\ref{fit20}), gives
\bea
\vec \Phi_{4,5}(m_{B_d},m_b) = 
\left(
\begin{array}{c}
 1.112(27) \\
 2.072(49) \\
\end{array}
\right)  \;,\hspace*{12mm} 
\vec \Phi_{4,5}(m_{B_s},m_b) = 
\left(
\begin{array}{c}
 1.121(12) \\
 2.101(35) \\
\end{array}
\right)  \;.
\eea
The two ways of reaching the physically relevant results 
are shown in fig.~\ref{fig44}. 
As in the previous subsection, for the final results 
we need to match back onto the QCD.
These values are presented in table~\ref{tab45}.
\begin{table}[h!] 
\begin{center} 
\begin{tabular}{|c|c|c||c|c|} \cline{2-5}
\multicolumn{1}{l|}{}&\multicolumn{2}{c||}
{\sl \small unconstrained}& 
\multicolumn{2}{c|}{\sl \small constrained} \\
\hline 
{\phantom{\Huge{l}}}\raisebox{-.2cm}{\phantom{\Huge{j}}}  
\hspace*{-5mm} {\sl \small Light quark} & 
$d$ & $s$  & $d$ & $s$ \\ \hline  
{\phantom{\Huge{l}}}\raisebox{-.2cm}{\phantom{\Huge{j}}}  
\hspace*{-5mm} $B_4^\msbar(m_b)$ & 
1.228(72) & 1.223(26) & 1.148(28)  & 1.157(13)\\ 
{\phantom{\Huge{l}}}\raisebox{-.2cm}{\phantom{\Huge{j}}}
\hspace*{-5mm} $B_5^\msbar (m_b)$  & 
1.784(90)& 1.798(50) & 1.724(41)  & 1.750(29)\\ 
 \hline
\end{tabular} 
\caption{\label{tab45}
\small{\sl Final results for the last two $B$-parameters defined in
eq.~(\ref{params}), in the  
$\msbar$(NDR) scheme of ref.~\cite{buras} at $\mu = m_b$. }}
\end{center}
\vspace*{-.3cm}
\end{table}
From tabs.~\ref{tab123} and \ref{tab45} we conclude that, 
for all the $B$-parameters, the extrapolations
 in which we do not include the static 
results lead to  higher values.

\section{Systematic uncertainties \label{sec4}}

The central values of this work are the ones obtained 
by combining the static HQET and QCD results, which are given in 
tabs.~\ref{tab123} and \ref{tab45}. 

We now need to attribute systematic errors to these results. We do not 
discuss the errors due to the use of the quenched approximation and refer 
to our results as quenched. The results of ref.~\cite{hashimoto}, however, 
are quite encouraging in that they indicate that the values of the 
parameters $B_{1,2,3}(m_b)$ remain practically unchanged after switching 
from $n_F=0$ to $n_F=2$.

\begin{itemize}

\item[$\circ$]  
Systematic uncertainties
present in the static results were estimated to be
in the range of $(3\div 4)\%$ for all the $\widetilde B$-parameters.
This error is almost entirely due to the choice of the
renormalization point $q^\ast$ at which we used the
boosted perturbative expressions. We varied $1/a \leq
q^\ast \leq \pi/a$, and then evolved the resulting
(continuum) B-parameters from $\mu = q^\ast$ to $\mu = m_b$.
The spread of values $\widetilde B(m_b)$ with respect to
the central one (obtained from $q^\ast = 2.6/a$),  has been
assigned to the systematic error. 

\item[$\circ$] The QCD values are obtained after the non-perturbative
renormalization in the RI/MOM scheme at $\mu = 2.8(1)$~GeV. We repeated the
whole procedure described in the previous section, but starting from 
our results obtained at 
$\mu = 1.9(1)$~GeV and at $\mu = 3.9(2)$~GeV.  
The final results get modified as follows:
\begin{itemize}
\item[--] $\Delta B_1/B_1^{\rm central} < \pm 1\%$;
\item[--] $\Delta B_2/B_2^{\rm central} \simeq \pm 1\%$;
\item[--] $\Delta B_3/B_3^{\rm central}\simeq \pm 8\%$;
\item[--] $\Delta B_4/B_4^{\rm central}\simeq -3 \%$;
\item[--] $\Delta B_5/B_5^{\rm central}\simeq +11 \%$.
\end{itemize}

\item[$\circ$]   Interpolation/extrapolation in the light quark mass is 
made linearly. 
For the average up/down quark mass we need to 
account for the possibility of the quadratic term in this extrapolation. 
As it can be seen from
fig.~\ref{fig2}, the extrapolations are smooth for all the bag parameters. 
If we include a quadratic term in the extrapolation, we obtain 
results which are fully compatible with the ones presented here.
This is true for both QCD and static HQET $B$-parameters.

\item[$\circ$] 
The value $a^{-1}(m_{K^\ast}) = 2.72(13)$~GeV, has been used throughout the 
paper.  
Another option would be to use the kaon decay constant, from which we obtain 
$a^{-1}(f_K) =  2.69(16)$~GeV. Being completely consistent with 
$a^{-1}(m_{K^\ast})$, this
choice affects our final results by only $+ 1\%$.

\item[$\circ$] Even though we use the improved action, our operators 
are not improved. Therefore, our results
for the bag parameters suffer from ${\cal O}(a)$ discretization errors. 
The hope is that these errors cancel in the ratios~(\ref{eq1},\ref{eq2}) 
from which the bag parameters are actually extracted. A conservative 
estimate on the size of these uncertainties can be obtained if we improve 
the axial current in the ratio $R_1$, as 
$A_0(t) \to A_0(t) + c_A (P_5(t+1) - P_5(t-1))/2$, with the known value 
for the parameter $c_A=-0.04$~\cite{alpha}. From this 
exercise we conclude the further increase in our final results for 
$B_1$ by $\sim 4\%$.

As for the HQET values, we checked that our values for the 
$\widetilde B_i$ (obtained at $\beta =6.0$) are indistinguishable from
the ones that can be extracted from the 
UKQCD data at $\beta =6.2$~\cite{ukqcd}. That gives us more confidence 
that the ${\cal O}(a)$ effects in the static HQET data are indeed small.

\item[$\circ$] We used the two-loop running coupling $\alpha_s(\mu)$ 
by taking $\Lambda_{\rm QCD}^{(n_F=0)}=0.25$~GeV. We tried to vary 
$\Lambda_{\rm QCD}^{(n_F=0)}$ by 10\% (which covers all the presently
available lattice estimates~\cite{heitger}), and see that the final results
vary in the range of $\pm 1.5$\%.   
\end{itemize}

We now write our results in a fully explicit form as:
\bea
B_1^{(d)\msbar} (m_b) = 0.87(4)(3)(0)\left({}^{+4}_{-2}\right) \;, &&
B_1^{(s)\msbar} (m_b) = 0.87(2)(3)(0)\left({}^{+4}_{-2}\right) \;,\cr 
&& \hfill \cr 
B_2^{(d)\msbar} (m_b) = 0.83(3)(3)(1)(2) \;,&&
B_2^{(s)\msbar} (m_b) = 0.84(2)(3)(1)(2) \;, \cr
&& \hfill \cr 
B_3^{(d)\msbar} (m_b) = 0.90(6)(3)(7)(2)\;, &&
B_3^{(s)\msbar} (m_b) = 0.91(3)(3)(7)(2)\;, \cr
&& \hfill \cr 
B_4^{(d)\msbar} (m_b) = 1.15(3)(4)\left({}^{+0}_{-4}\right)(3)\;,  &&
B_4^{(s)\msbar} (m_b) = 1.16(2)(4)\left({}^{+0}_{-4}\right)(3)\;,  \cr
&& \hfill \cr 
B_5^{(d)\msbar} (m_b) = 1.72(4)(5)\left({}^{+19}_{-00}\right)(3) \;,&& 
B_5^{(s)\msbar} (m_b) = 1.75(3)(5)\left({}^{+20}_{-00}\right)(3) \;, 
\eea
where, besides the first statistical errors, 
the following sources of the systematic 
uncertainty are being written out respectively: 
systematics of the calculation in the static limit of
HQET, the error in the renormalization of $B$-parameters computed in
QCD, combined error due to the variation of $a^{-1}$ and of 
$\Lambda_{\rm QCD}^{(n_F=0)}$ (and also due to the improvement of 
the axial current in the case of $B_1$). 
After adding all  systematic errors in squares we arrive at the 
complete set of results already given in table~\ref{tab0}.

To be able to fully reconstruct the numbers that we presented in 
table~\ref{tab0},
we also need to provide the reader with the formulae allowing the conversion  
of the parameters  $B_2(m_b)$ and $B_3(m_b)$ from the  
$\msbar$(NDR) scheme of ref.~\cite{beneke} to the one 
of ref.~\cite{buras}. 
This is achieved by using the following formula
\bea
&&
\left(
\begin{array}{c}
\langle O_2(\mu)\rangle  \\
\langle O_3(\mu)\rangle  \\
\end{array}
\right)^{\msbar\ \mbox{\cite{buras}}} =  
\left[\; \mathbb{I} \ +\  {\alpha_s(\mu)\over 12 \ \pi} \ \left( 
\begin{array}{cc}
  -11 & 1  \\
 1 &  5   \\
\end{array} \right)\; \right] \ 
\left(
\begin{array}{c}
\langle O_2(\mu)\rangle  \\
\langle O_3(\mu)\rangle  \\
\end{array}
\right)^{\msbar\ \mbox{\cite{beneke}}} \;,
\eea
which we obtained after rotating the operators $Q^{SLL}_{1,2}(\mu)_{\msbar}$
of ref.~\cite{buras} to the SUSY basis~(\ref{basisc}).

\section{Concluding remarks \label{sec5}}

In this paper we computed the $B$-parameters for all five $\Delta B=2$ operators.
The extrapolation of the results obtained directly in lattice QCD in the 
region of masses $m_P \sim 2$~GeV to the physically interesting mass
$m_{B_{d/s}}$, has been constrained by using the static HQET result. 
The matching QCD $\leftrightarrow$ HQET and running in each of the two theories 
have been made by the consistent use of the perturbative expressions known at NLO.
The final results are presented in three renormalization
schemes (see table~\ref{tab0}).

Our results can be improved in many ways. We combined the results of the QCD 
lattice simulations performed at $\beta=6.2$ with the HQET ones obtained at $\beta=6.0$. 
Naturally, a good strategy would be to do the computation at the same value of 
$\beta$ in both theories, to vary the value of $\beta$ ({\it i.e.} of 
the lattice spacing) and attempt extrapolating to the continuum limit.
All numbers are obtained in the quenched approximation ($n_F=0$). An investigation of 
the sea quark effects on our quenched values  by repeating the analysis
performed in this paper with $n_F=2$, would be very welcome.

\subsection*{Acknowledgements}

The work of three of us (V.G., G.M., J.R) has been supported by the 
European Community's Human potential programme under HPRN-CT-2000-00145
Hadrons/LatticeQCD. 
V.G. and J.R. have been supported in part by CICYT under Grant AEN-96-1718, 
by DGESIC under the Grant PB97-1261 and by the Generalitat Valenciana 
under the Grant GV98-01-80. J.R. thanks the Departamento de 
Educaci\'on of the Gobierno Vasco for a predoctoral fellowship.
D.B. thanks the University ``La Sapienza" for support. 
\newpage
\subsection*{Appendix: Non-perturbative calculation of the 
renormalization and subtraction constants in the (Landau)RI/MOM scheme\label{app1}}

In this appendix we give the numerical values for the resulting matrices 
of the renormalization and subtraction constants which are obtained in the 
(Landau)RI/MOM scheme by using the method of refs.~\cite{npr,DI32}. 
These are computed in the following basis of operators:
\bea \label{baseF}
&&Q_1 = 
\overline q^i \ga_\mu (1 - \ga_5) q^i\ \overline q^j  \ga_\mu (1 - \ga_5)q^j 
\;, \nonumber \\
&&Q_2 = 
\overline q^i \ga_\mu (1 - \ga_5) q^i\ \overline q^j  \ga_\mu (1 + \ga_5)q^j 
\;, \nonumber \\
&&Q_3 =  
\overline q^i  (1 + \ga_5) q^i\ \overline q^j  (1 - \ga_5)q^j 
\;,  \\
&&Q_4 =  
\overline q^i  (1 - \ga_5) q^i\ \overline q^j  (1 - \ga_5)q^j 
\;, \nonumber \\
&&Q_5 = {1\over 2}   
\overline q^i \sigma_{\mu \nu} (1 - \gamma_5) q^i\ \overline q^j   
\sigma_{\mu \nu} (1 - \gamma_5) q^j \quad (\mu > \nu)
\;, \nonumber 
\eea
which are equivalent to those appearing in eq.~(\ref{basisc}) (after setting $\bar b\to \bar q$):
\bea
&&Q_1 = O_1\;,\hspace*{11mm} Q_2 = -2 O_5\;,\cr
&&\cr
&& Q_3 = O_4\;,\hspace*{11mm} Q_4 = O_2\;, \cr
&&\cr
&&Q_5 =O_2 + 2 O_3\;.
\eea
The difference between these and the results for the renormalization constants presented in
refs.~\cite{ape1,ape2} is that the present renormalization and subtraction constants 
are not polluted by the Goldstone 
boson contributions. To eliminate those, we applied the recipe of ref.~\cite{DI32}.
Since that paper has not  been released yet, we briefly explain the
main steps here.
\begin{itemize}
\item[$\odot$] Starting from the 4-quark Green functions computed in the Landau gauge, 
with all momenta in the external legs equal, 
$G_i(p) = \langle q(p)\ \bar q(p)\ Q_i\  \bar q(p)\ q(p)\rangle$, one  
constructs the amputated ones as
\bea
\Lambda_i(p) = \left( \prod_{k=1}^{4} S^{-1}(p)\right) G_i(p)\;,
\eea
where $S^{-1}(p)$ stands for the inverse quark propagator.
\item[$\odot$]
The amputated Green functions are projected onto various Dirac 
structures as
\bea \label{proj}
\Bigl( \Gamma_i(p) \Bigr)_{j} = {\rm Tr}\Bigl[ 
 \Lambda_i(p) P_j \Bigr]\,,
\eea 
where $P_j$ are suitable projectors satisfying the 
orthogonality relation
\bea \label{ort}
 \Bigl( \Gamma_i^{(0)}(p) \Bigr)_{j} = {\rm Tr}\Bigl[ 
 \Lambda_i^{(0)}(p) P_j \Bigr] = \delta_{ij}\; ,
\eea
where $\Lambda_i^{(0)}(p)$ stands for the tree level amputated Green functions.
The explicit expressions 
for the projectors $P_j$ can be found in ref.~\cite{npr} (eq.~(37)).
\item[$\odot$] Eq.~(\ref{ort}) is turned into the RI/MOM renormalization condition
as
\bea \label{rcmom}
\Biggl. \Bigl( \hat \Gamma_i(p/\mu) \Bigr)_{j}\Biggr|_{p^2=\mu^2} = 
\delta_{ij}\,,
\eea
where the renormalized amputated Green function, $\hat \Gamma_i(p/\mu)$, 
is expressed as
\bea 
\hat \Gamma_i(p/\mu) =  \Gamma_k(p)\biggl( \delta_{kj} 
+ \Delta_{kj}(g_0^2) \biggr) Z_{ji} (\mu,g_0^2)\,,
\eea
up to an overall wave function renormalization, which is trivial to compute  
after imposing the vector Ward identity on the quark propagator.
The condition~(\ref{rcmom}) is applicable for virtualities 
$\Lambda^2_{\rm QCD} \ll p^2 \ll (\pi/a)^2$.

\item[$\odot$]
Thus, for each of the operators from the basis~(\ref{baseF}), 
one obtains 5 equations from which the subtraction ($\Delta(g_0^2)$)
and renormalization ($Z(\mu;g^2_0)$) constants are computed. In 
matrix form, the final result writes
\bea
\vec Q(\mu)\ =\ Z(\mu;g_0^2)\  
\biggl[ \mathbb{I} + \Delta(g_0^2) \biggr] \ 
\vec Q(g_0^2)\;.
\eea
The structure of the $Z$-matrix is determined by the chirality, {\it i.e.}
$Q_1(\mu)$ does not mix with any other operator,~\footnote{
In other words $Z_{11}(\mu) \neq 0$, whereas  
$Z_{12} = Z_{13} =Z_{14} =Z_{15} =0$.} $Q_{2,3}(\mu)$ 
mix with each other but not with the other operators. The same 
goes for the $Q_{4,5}(\mu)$ operators. The remaining elements
of the $5\times 5$ matrix are filled by the subtraction constants
$\Delta_{ij}(g_0^2)$.
\item[$\odot$]
In practice, however, the above procedure is implemented by 
computing $\hat \Gamma_i(p,\kappa_q)$ at several values of the (light)
quark mass ({\it i.e.} various $\kappa_q$), 
followed 
by the extrapolations of each $\Delta_{ij}(g_0^2)$ and 
$Z_{ij}(\mu, g_0^2)$ to the chiral limit. 
This extrapolation can be dangerous 
because the operators are inserted at zero momentum (all external 
legs in the Green function have the same momentum), and the 
coupling to the Goldstone boson contaminates the short
distance behaviour (which we are interested in). 
In particular, we find that for the vertices of the structure 
$\gamma_5\otimes \gamma_5$ this coupling is indeed large. 
In addition, via projections~(\ref{proj}), it may pollute the
extraction of the renormalization and subtraction constants for the 
other operators. Therefore, for each projected amputated four-quark
Green function, one should subtract the Goldstone contribution. 
For the parity even operators  
(which are the ones that we consider in this paper), 
this Goldstone contribution can appear as a pole, 
but also as a double pole, {\it i.e.}:
\bea
 \Bigl( \Gamma_i(p;\kappa_q) \Bigr)_{j} \equiv \Gamma_{ij}(p;\kappa_q) 
= \alpha_{ij}(p) \ + \ {\beta_{ij}(p) \over m_{P}^2}  \ + 
\ {\gamma_{ij}(p) \over m_{P}^4}  \ + \ \delta_{ij} \ m_{P}^2.
\eea
Note that we also added a term $\delta \ m_{P}^2$, to account for the 
the linear dependence in the quark mass ($m_{P}^2 \propto m_q$).~\footnote{
The linear dependence in the quark mass arises after the cancellation 
of the quadratic quark mass term against the Goldstone 
pole contribution. More detailed discussion will be presented in ref.~\cite{DI32}.}  
A judicious way to subtract the Goldstone contributions,  
and thus to reach the term $\alpha_{ij}(p)$,
is to consider the following combination for the fit with the data~\cite{DI32}
\bea \label{subtr}
{ m_{P_1}^2 \Gamma_{ij}(p;\kappa_{q_1}) - m_{P_2}^2 \Gamma_{ij}(p;\kappa_{q_2}) 
\over m_{P_1}^2 - m_{P_2}^2 } \ =\  
\alpha_{ij}(p) \ - \ {\gamma_{ij}(p) \over m_{P_1}^2  m_{P_2}^2}  \ + \ 
{\delta_{ij}(p)} \Bigl( m_{P_1}^2 + m_{P_2}^2 \Bigr)  
\eea
by which the pole-like contribution is automatically eliminated. 
This is to be performed for each value of $p$ and accounting for 
all the mass combinations.~\footnote{ Besides 
the light masses already mentioned in the
previous section, for the discussion of the renormalization constants 
we also had the results for $\kappa_q = 0.1333$ at our disposal. In 
this way, we could make 6 combinations 
and therefore a safe extrapolation to the chiral limit.} The resulting  
$\alpha_{ij}(p)$ is thus the chiral value of the projection $\Gamma_{ij}(p)$ 
 free from the Goldstone boson contamination.
\end{itemize}

The procedure sketched above has been applied at three values of 
the renormalization scale: $a \mu = 0.71$, $1.03$ 
and $1.41$. In physical units, these values correspond to 
$\mu = 1.9(1)$~GeV, $2.8(1)$~GeV and $3.9(2)$~GeV, respectively. 
The complete list of results for the subtraction ($\Delta_{ij} (a)$) and 
renormalization constants ($Z_{ij} (\mu a)$) is presented in
table~\ref{tabREN}.

As for the renormalization constants for the  bilinear quark operators, 
which  are
necessary to compute the ratios~(\ref{eq1}) and (\ref{eq2}), we use the 
following values~\cite{charm}:
\bea
Z_P^{\rm RI/MOM} \Bigl(1.9(1)\ \gev\Bigr)&=&0.510(9)\,;\nonumber \\
Z_P^{\rm RI/MOM} \biggl(2.8(1)\ \gev\biggr)&=&0.575(7)\,;\\
Z_P^{\rm RI/MOM} \biggl(3.9(2)\ \gev\biggr)&=&0.630(6)\,.\nonumber
\eea
In addition, in the same study, we obtained $Z_A = 0.814(4)$. Notice that 
we used the method of ref.~\cite{leo}, to avoid the large Goldstone boson 
contribution to the value of $Z_P^{\rm RI/MOM}$~\cite{alain}. As for the 
constant $Z_A$, our value is consistent with the findings of other lattice 
groups~\cite{alpha}. 

\newpage
\begin{table}[h!] 
\begin{center} 
\begin{tabular}{|c|c|c|c|} 
\hline 
{\phantom{\Huge{l}}}\raisebox{-.2cm}{\phantom{\Huge{j}}}
\hspace*{-7mm}{\sl Scale ($\mu$)}  & {1.9(1) GeV} & 2.8(1) GeV   &  3.9(2) GeV 
          \\ \hline \hline 
{\phantom{\Huge{l}}}\raisebox{-.2cm}{\phantom{\Huge{j}}}  
\hspace*{-4mm} $Z_{11}(\mu)$ 
& 0.663(6)  & 0.645(9)   & 0.644(13)    \\ \hline
 $\Delta_{12}$ 
& -0.072(3)    &       -0.069(11)   &	-0.063(10)     \\ 
 $\Delta_{13}$ 
& -0.015(2)    &	     -0.011(4)    &	-0.014(2)     \\ 
 $\Delta_{14}$ 
& 0.020(2)     &	      0.021(7)    &	 0.015(6)     \\ 
 $\Delta_{15}$ 
& 0.011(3)     &	      0.006(1)    &	 0.005(1)      \\  \hline \hline
{\phantom{\Huge{l}}}\raisebox{-.2cm}{\phantom{\Huge{j}}}
\hspace*{-4mm} $Z_{22}(\mu)$ 
 &  0.723(5) & 0.691(7) & 0.683(5)   \\ 
{\phantom{\Huge{l}}}\raisebox{-.2cm}{\phantom{\Huge{j}}}
\hspace*{-4mm} $Z_{23}(\mu)$ 
& 0.315(3)  & 0.257(7)   & 0.202(9)  \\ \hline
 $\Delta_{21}$ 
& -0.052(2)    &	     -0.055(9)    &	-0.050(9)     \\ 
 $\Delta_{24}$ 
& -0.250(7)    &	     -0.169(20)   &	-0.168(19)     \\ 
$\Delta_{25}$ 
& 0.013(2)     &	      0.014(2)    &	 0.015(1)   \\ \hline\hline
{\phantom{\Huge{l}}}\raisebox{-.2cm}{\phantom{\Huge{j}}}
\hspace*{-4mm} $Z_{32}(\mu)$ 
& 0.023(1)  & 0.022(1)    & 0.021(3)  \\ 
{\phantom{\Huge{l}}}\raisebox{-.2cm}{\phantom{\Huge{j}}}
\hspace*{-4mm} $Z_{33}(\mu)$ 
& 0.322(12)  &  0.392(20)  & 0.467(13) \\ \hline
 $\Delta_{31}$ 
& 0.018(1)     &	      0.018(4)    &	 0.014(3)    \\ 
 $\Delta_{34}$ 
& 0.351(10)    &	      0.233(34)   &	 0.220(31)   \\ 
 $\Delta_{35}$ 
& -0.008(1)    &	     -0.007(3)    &	-0.005(1)  \\ \hline\hline
{\phantom{\Huge{l}}}\raisebox{-.2cm}{\phantom{\Huge{j}}}
\hspace*{-4mm} $Z_{44}(\mu)$ 
 & 0.414(9)  &  0.473(17)   &  0.534(13) \\ 
{\phantom{\Huge{l}}}\raisebox{-.2cm}{\phantom{\Huge{j}}}
\hspace*{-4mm} $Z_{45}(\mu)$ 
& -0.017(2)  & -0.015(2)     &  -0.015(4) \\ \hline
$\Delta_{41}$ 
& 0.008(1)     &	      0.008(3)    &	 0.005(1)    \\ 
$\Delta_{42}$ 
& 0.009(1)     &	      0.001(2)    &	 0.001(1)   \\ 
$\Delta_{43}$ 
& 0.208(4)     &	      0.143(16)   &	 0.144(15)  \\ \hline\hline
{\phantom{\Huge{l}}}\raisebox{-.2cm}{\phantom{\Huge{j}}}
\hspace*{-4mm} $Z_{54}(\mu)$ 
 &-0.307(4)   & -0.233(7)   & -0.176(7) \\ 
{\phantom{\Huge{l}}}\raisebox{-.2cm}{\phantom{\Huge{j}}}
\hspace*{-4mm} $Z_{55}(\mu)$ 
&  0.914(9) & 0.814(16) & 0.761(18)  \\ \hline
$\Delta_{51}$ 
& 0.009(1)     &	      0.008(2)    &	 0.005(1)   \\ 
$\Delta_{52}$ 
& 0.009(1)     &	      0.005(3)    &	 0.008(1)  \\ 
$\Delta_{53}$ 
& 0.121(1)     &	      0.084(6)    &	 0.089(8)    \\ 
 \hline  
\end{tabular} 
\caption{\label{tabREN}
\small{\sl The values of the renormalization $Z_{ij}(\mu)$ and subtraction 
constants $\Delta_{ij}$
computed non-perturbatively in the (Landau) RI/MOM scheme at three different
values of the renormalization scale $\mu$ at $\beta = 6.2$. 
}}
\end{center}
\end{table}

\newpage


\end{document}